\documentclass[10pt,twocolumn,aps,showpacs,nofootinbib,floatfix,nobibnotes]{revtex4-2}  
\usepackage[utf8]{inputenc}  
\usepackage[english]{babel}  
\usepackage{amsmath, bm, amssymb}  
\usepackage{mathtools}  % \coloneqq, \xRightarrow  
\usepackage{dsfont}     % math symbol (\mathds{1})  
\usepackage{graphicx}  
\usepackage{tikz}  
\usetikzlibrary{shapes,arrows}  
\usepackage{enumitem}  
\usepackage[colorlinks=true,linkcolor=blue,citecolor=blue,urlcolor=blue]{hyperref}  
\makeatletter  
\let\label\ltx@label   
\makeatother  
\usepackage{booktabs} 
\usepackage{makecell}
\usepackage[normalem]{ulem}
\usepackage[utf8]{inputenc}
\usepackage{textcomp}
\usepackage{tikz}
\usetikzlibrary{arrows.meta}
\usepackage{listings}% http://ctan.org/pkg/listings
\begin{document}

\title{Efficient mapping of multi-constraint satisfaction problems to Rydberg platforms}

\author{Robert Gloeckner$^{1}$}
\email{robert.gloeckner@lhind.dlh.de}
\author{Shahram Panahiyan$^{2,3}$ }
\email{shahram.panahiyan@mpsd.mpg.de}
\author{Frederik Koch$^{2}$}
\author{Dieter Jaksch$^{2,4,5}$}
\author{Joseph Doetsch$^{1}$}

\affiliation{$^1$ Lufthansa Industry Solutions, Südportal 7, 22848 Norderstedt, Germany}
\affiliation{$^2$ University of Hamburg, Luruper Chaussee 149, 22761 Hamburg, Germany}
\affiliation{$^3$  Max Planck Institute for the Structure and Dynamics of Matter, Luruper Chaussee 149, 22761 Hamburg, Germany}
\affiliation{$^4$ The Hamburg Centre for Ultrafast Imaging, Luruper Chaussee 149, Hamburg D-22761, Germany}
\affiliation{$^5$ Clarendon Laboratory, University of Oxford, Parks Road, Oxford OX1 3PU, United Kingdom}

\date{\today}

\begin{abstract}

We present a hardware-native gadget framework for solving constraint satisfaction problems on Rydberg quantum computing architectures.
Our approach introduces a compact $xor_1$ gadget that enforces exactly-one constraints, ubiquitous in combinatorial optimization, directly through geometric embedding and blockade interactions.
A key advantage of the $xor_1$ gadget is its fixed, problem-size-independent detuning requirements: enforcing constraints through blockade interactions eliminates the need for large penalty terms, thereby substantially reducing the detuning range compared to Quadratic Unconstrained Binary Optimization (QUBO) formulations and improving experimental feasibility.
By tailoring the construction to the geometric connectivity of Rydberg atom arrays, the framework bypasses the all-to-all physical couplings often assumed in logical encodings. This enables embeddings compatible with planar layouts and avoids highly connected arrangements.
We develop scalable implementations that reduce atom count and connectivity overhead while avoiding extensive classical preprocessing, making them compatible with near-term neutral-atom hardware.
As illustrations, we apply our framework to the gate-assignment and $N$-queens problems, highlighting its practicality, resource efficiency, and hardware compatibility. In these examples, we observe reductions in detuning range of up to $99\%$ and savings in atom count and connectivity overhead of up to $54\%$ compared to the QUBO method.
These results establish a route toward implementing large-scale combinatorial optimization on Rydberg platforms beyond the limits of existing encodings.
\end{abstract}

\maketitle

\section{Introduction}

%Optimization satisfaction is a central challenge in many industrial and engineering applications, including portfolio optimization, logistics, scheduling, and design problems \cite{marzecPortfolioOptimizationApplications2016, crainicCombinatorialOptimizationApplications2024, prataOverviewIndustrialEngineering2025}. Such tasks typically involve searching large, complex, and often non-convex or constrained solution spaces, where classical heuristics, gradient-based methods, and mathematical programming approaches face severe computational limitations as problem size increases \cite{crescenziCompendiumNPOptimization1995}. In many cases, the exponential growth of the solution space renders the identification of optimal or near-optimal solutions computationally intractable, even with advanced relaxation and approximation techniques \cite{burerNonconvexMixedintegerNonlinear2012, floudasGlobalOptimization21st2005, xuRelaxationsBinaryPolynomial2024}.

Optimization and constraint satisfaction problems (CSP) are central challenges in many industrial and engineering applications, including portfolio optimization, logistics, scheduling, and design problems \cite{marzecPortfolioOptimizationApplications2016, crainicCombinatorialOptimizationApplications2024, prataOverviewIndustrialEngineering2025}. Such tasks typically involve searching large, complex, and often non-convex or constrained solution spaces, where classical heuristics, gradient-based methods, and mathematical programming approaches can face significant computational limitations as problem size increases \cite{crescenziCompendiumNPOptimization1995}. In many cases, the exponential growth of the solution space renders the identification of feasible or optimal solutions computationally intractable, even with advanced relaxation and approximation techniques \cite{burerNonconvexMixedintegerNonlinear2012, floudasGlobalOptimization21st2005, xuRelaxationsBinaryPolynomial2024}.

Emerging quantum computing technologies offer a new paradigm for tackling such problems \cite{abbasChallengesOpportunitiesQuantum2024, volpeImprovingSolvingOptimization2025, grangeIntroductionVariationalQuantum2024a}. By exploiting quantum superposition and entanglement, quantum algorithms may enable new strategies for exploring complex optimization landscapes beyond classical capabilities \cite{farhi2014quantumapproximateoptimizationalgorithm, abbasChallengesOpportunitiesQuantum2024,koch2025resourceefficientquantumoptimizationhigherorder}. As quantum hardware advances, integrating quantum methods into optimization workflows could accelerate solutions to previously intractable problems across scientific and industrial domains \cite{cerezo2021variational,sachdevaQuantumOptimizationUsing2024, quintonQuantumAnnealingApplications2025, soodIndustrialProgressQuantum2025}.

%A particularly promising direction within this landscape is the use of controlled quantum many-body dynamics as a computational resource. The principal idea is to steer the system’s evolution such that its final state encodes the solution to a given optimization problem, effectively transforming quantum dynamics into a computational resource. Rydberg-atom quantum computing platforms provide a natural setting for this approach: owing to the Rydberg blockade mechanism, they directly implement unit-disk graph interactions, allowing the maximum-weighted independent-set problem to be embedded into the native many-body Hamiltonian. 

A particularly promising direction within this landscape is the use of controlled quantum many-body dynamics as a computational resource \cite{Ebadi2021,browaeys2020,Scholl2021,r54t-myhc,Sylvain2019,dtlf-2q82,buchler2025quantumdoublessymmetricblockade,Steinert2023,liu2025timeserieslearningmanybody,hudomal2025ergodicitybreakingmeetscriticality,büchler2025quantumdoublessymmetricblockade,martins2026quantumsimulationrydbergions,kokail2026inversequantumsimulationquantum}. In this paradigm, the evolution of a quantum system is engineered such that its final state encodes the solution to a target optimization problem or CSP. Rydberg-atom quantum computing platforms are especially well suited to this approach: the Rydberg blockade mechanism naturally implements interactions on unit-disk graphs (UDGs), enabling the direct embedding of maximum-weighted independent set (MWIS) problems into the native many-body Hamiltonian \cite{pichler2018, Ebadi2022,Lanthaler2023}. This native correspondence between physical interactions and graph-theoretic constraints distinguishes Rydberg arrays from more abstract gate-based optimization implementations.

%Building on this observation, Nguyen \textit{et al.} introduced a family of gadget constructions that significantly broaden the class of optimization problems realizable on Rydberg atom arrays \cite{nguyen2023quantum}. While powerful, the current implementations face significant practical challenges. Encoding general problem instances often incurs quadratic overhead in system size and requires substantial classical preprocessing. In addition, large detunings are typically employed to enforce energetic constraints, leading to increased laser-power requirements, enhanced sensitivity to amplitude and phase noise, and reduced coherence times \cite{Ebadi2021,Scholl2021,Oliveira2025,Bombieri2025,Schuetz2025}. Together, these factors limit scalability and degrade the fidelity of the effective optimization landscape.

Building on this observation, Nguyen et al. introduced a family of gadget constructions that significantly broaden the class of combinatorial optimization problems and CSPs realizable on Rydberg atom arrays (RAAs) \cite{nguyen2023quantum}. While powerful, existing implementations face important practical challenges. QUBO-based formulations typically enforce hard constraints through large energetic penalties, leading to significant overhead in system size and requiring substantial classical preprocessing to generate the corresponding embeddings. 
Moreover, the large detunings required to impose such penalties consume a substantial fraction of the finite detuning range available in current Rydberg platforms. 
%This, in turn, compresses the dynamic range remaining for encoding the cost function, reducing spectral separation between feasible configurations and increasing sensitivity to noise and control imperfections \cite{Ebadi2021,Scholl2021,Oliveira2025,Bombieri2025,Schuetz2025}. Together, these effects limit scalability and degrade the fidelity of the effective optimization landscape.
As a result, the available dynamic range must be shared between enforcing constraints and encoding the cost function, limiting the energy scale separation between feasible and infeasible configurations as well as between competing feasible solutions. This compression reduces spectral distinguishability, increasing sensitivity to noise and control imperfections \cite{Ebadi2021,Scholl2021,Oliveira2025,Bombieri2025,Schuetz2025}.
Together, these effects hinder both the faithful realization of constraints and the accurate representation of the optimization landscape, ultimately limiting scalability and solution quality.

%In this work, we introduce a gadget-based framework that addresses these limitations by reformulating constraint optimization problems using clique-based constructions that map efficiently onto Rydberg architectures. Central to our approach is a compact $xor_1$ gadget that enforces exactly-one selection rules, a ubiquitous constraint in combinatorial optimization problems. By expressing these problems as MWIS instances on UDGs, our formulation maintains a direct correspondence between graph nodes and individual Rydberg atoms, with blockade-induced edges encoding the requisite exclusivity constraints (see Fig. \ref{fig:G1}). This mapping effectively avoids extensive classical preprocessing and substantially reduces the required atom count and connectivity overhead compared to previous gadget-based approaches developed in Ref. \cite{nguyen2023quantum} (see Table \ref{tab:xorbenchmark}). We further demonstrate scalable implementations of the $xor_1$ gadget that operate with reduced detuning requirements, alleviating experimental constraints. To illustrate the practical advantages of our method, we apply our framework to two classes of NP-hard problems, namely, the gate assignment and $N$-queens problems, illustrating improved resource efficiency and compatibility with current state-of-the-art neutral-atom hardware for quantum optimization.

In this work, we introduce a gadget-based framework, the \(xor_1\) gadget, that addresses these limitations by reformulating constraint satisfaction problems using clique-based constructions that map efficiently onto Rydberg architectures. Our primary focus is on solving CSPs, while we also briefly discuss extending the framework to optimization which is conceptually straightforward.
The paper is organized as follows. Section \ref{sec:main} presents the main results. Section \ref{sec:math} details the mathematical description of the \(xor_1\) formulation, including the clique graph representation, its Boolean optimization reformulation, and multi-constraint reduction via the \(xor_1\) encoding. Section \ref{sec:cliquetoRyd} covers the mapping of cliques to unit-disk graphs and Rydberg platforms. Section \ref{sec:Xor1Alll} introduces the \(xor_1\) gadget and composite structures, including gadget construction, utilization of crossing and copy gadgets, MWIS analysis, realization of conjunctions, scaling analysis, and generalization from constraint satisfaction to optimization. Section \ref{sec:gate} presents a minimal example of gate assignment. Section \ref{sec:conclusion} concludes, with the Appendix providing larger gate-assignment illustrations (\ref{sec:gateadd}) and the \(N\)-queens problem (\ref{sec:nqueens}) realization via \(xor_1\) gadget.

\section{Main results} \label{sec:main}

\begin{figure*}[htbp]
    \centering
    \includegraphics[width=1\linewidth]{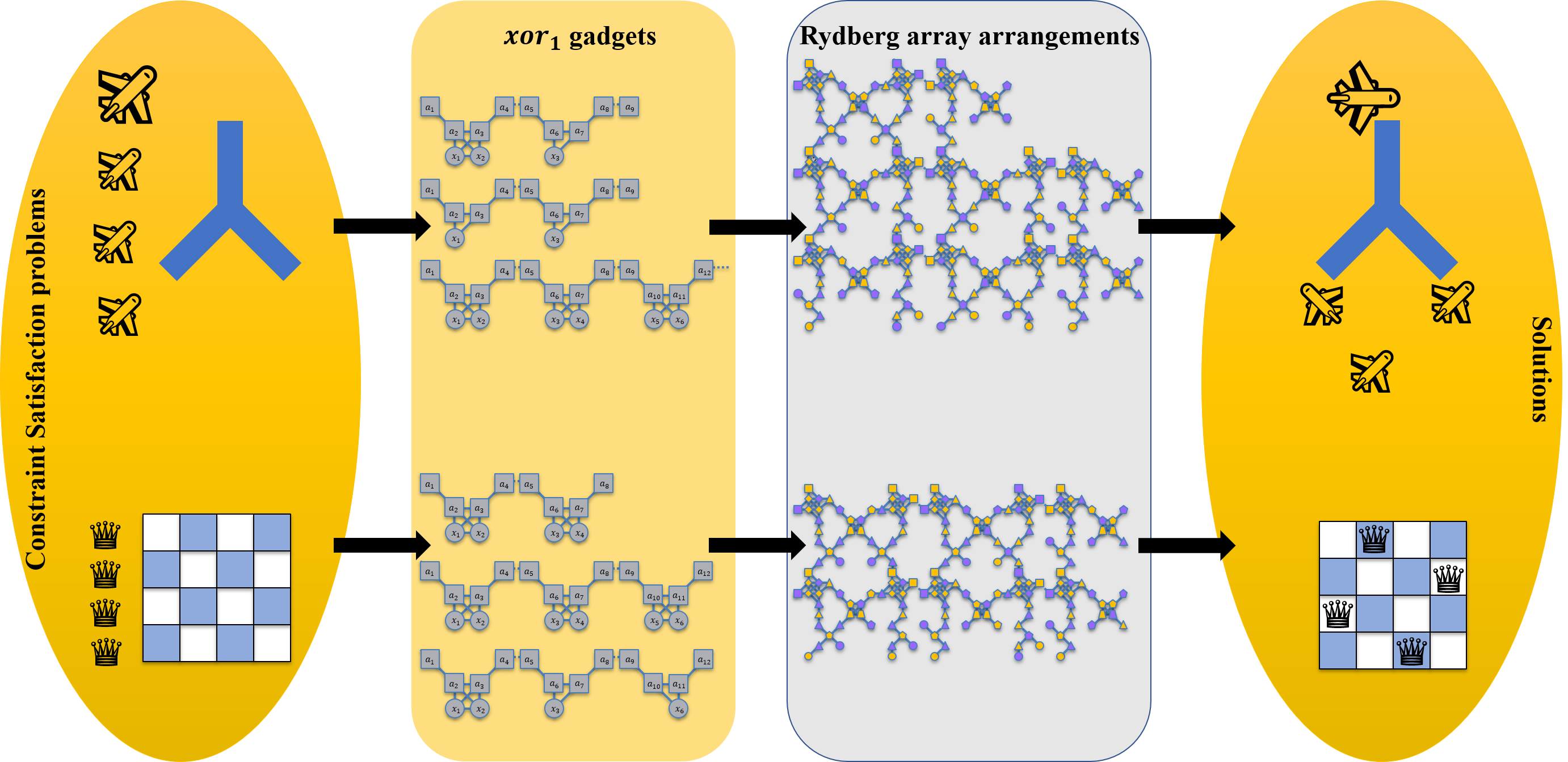}    
    \caption{Schematic representation of the application of the \(xor_1\) gadget developed in this work for solving CSPs. Different classes of CSPs, such as the airport gate-assignment problem and the \(N\)-queens problem, can be addressed within this framework. In the first step, problem-specific constraint structures are constructed using the \(xor_1\) gadget. These structures are then embedded onto the RAA using composite structures. 
    %shown in Figs.~\ref{fig:Cross-Copy} and \ref{fig:additional}. 
    The ground state of the resulting RAA encodes solutions to the corresponding CSPs.
 }
    \label{fig:G1}
\end{figure*}

The central ideas of this work are summarized schematically in Fig.~\ref{fig:G1}. We introduce a $xor_1$ gadget that enables the solution of binary optimization problems and CSPs with exactly-one and zero-or-one constraints within a clique-based incompatibility-graph formalism (Sec.~\ref{sec:math}). In this framework, cliques represent collections of mutually exclusive assignments that enforce selection constraints. These clique-based constructions can be embedded as subgraphs of UDGs, whose MWIS formulation maps directly to problem instances realizable on RAAs (Sec.~\ref{sec:cliquetoRyd}).

Leveraging this correspondence, we design geometric $xor_1$ gadgets that encode exactly-one and zero-or-one constraints through blockade interactions and Rydberg atom arrangements rather than energetic penalty terms.
%The gadget structures are shown in Figs.~\ref{fig:XOR}, \ref{fig:Cross-Copy}, and \ref{fig:additional}, with construction details provided in Sec.~\ref{sec:Xor1Alll}. 
The advantages of the proposed $xor_1$ gadget over existing approaches are fourfold.

%First, although clique formulations require all-to-all connectivity among optimization-variable vertices, such connectivity need not be physically realized on the Rydberg platform. The (xor_1) gadget enables embedding arbitrary incompatibility (MWIS) graphs into geometrically feasible atom arrangements while preserving clique constraints. As a result, large logical cliques can be implemented without long-range interactions or complex geometries, with effective connectivity enforced through blockade-mediated gadget structure.

\begin{table}
\centering
\begin{tabular}{l|c | c | c}
\toprule
\textbf{} & \textbf{Example I} & \textbf{Example II} & \textbf{Example III} \\
\midrule
\textbf{gates} & 3 &  3 & 5 \\
\textbf{slots} & 7 & 10 & 15 \\
\textbf{flights} & 5 & 10 & 20 \\
\textbf{opt. var.} & 13 & 26 & 81 \\
%\textbf{objective} & 4238 & 13126 & 94341 \\
\textbf{\(xor_1\)-detuning} & 6 & 6 & 6 \\
\textbf{QUBO-detuning} & 416 & 1320 & 9156 \\
\textbf{\(xor_1\)-atoms} & 2693 & 8119  & 57124 \\
\textbf{QUBO-atoms} & 3076 & 10298 & 89696 \\
\bottomrule
\end{tabular}
\caption{
%Comparison of three gate-assignment example scenarios using the proposed $xor_1$ gadget in this work and the QUBO-based method introduced in Ref.~\cite{nguyen2023quantum}. The increasing number of optimization variables (opt. var.) reflects the growing number of flights to be assigned to gates under time-slot and geometric restrictions (see Sec.~\ref{sec:gateadd} for details). The $xor_1$- and QUBO-detuning correspond to the maximum range of Rydberg atom detuning needed for the array to enforce feasible solutions/satisfy all constraints. The quantities labeled $xor_1$-atoms and QUBO-atoms denote the number of Rydberg atoms required to implement the corresponding Rydberg atom array encodings. 
Comparison of three gate-assignment example scenarios using the proposed $xor_1$ gadget in this work and the QUBO-based method introduced in Ref.~\cite{nguyen2023quantum}. The increasing number of optimization variables (opt. var.) reflects the growing number of flights to be assigned to gates under time-slot and geometric restrictions (see Sec.~\ref{sec:gateadd} for details). The $xor_1$- and QUBO-detuning correspond to the maximum range of Rydberg atom detunings required for the array to enforce feasible solutions and satisfy all constraints. 
Detuning values are reported in dimensionless units normalized by the characteristic Rabi frequency $\Omega_0$ \cite{nguyen2023quantum}, which we set to $\Omega_0 = 1$ for convenience. 
The quantities labeled $xor_1$-atoms and QUBO-atoms denote the number of Rydberg atoms required to implement the corresponding RAA encodings.
}
\label{tab:xorbenchmark}
\end{table}

%First, although cliques in clique-based formulations require all-to-all connectivity among the vertices associated with optimization variables, such connectivity need not be physically realized on the Rydberg platform. Our gadget circumvents this limitation by enabling the embedding of arbitrary incompatibility (MWIS) graphs into geometrically feasible Rydberg atom arrangements while preserving the clique constraints. Consequently, large logical cliques can be implemented without resorting to long-range interactions or otherwise complex geometries. Instead, the effective all-to-all connectivity is mediated through blockade interactions and the structure of the gadget itself, so that the required logical connectivity is enforced at the level of the construction rather than through direct physical coupling.

First, the detuning requirements of the $xor_1$ gadget are fixed and independent of problem size, unlike penalty-based encodings such as Ref.~\cite{nguyen2023quantum}. Because constraints are implemented natively through Rydberg blockade interactions rather than energetic penalties, large problem-dependent detunings are avoided. This preserves the available detuning range for encoding the cost function and maintains spectral separation between feasible configurations, improving robustness of the effective optimization landscape. Consequently, the approach supports scaling to larger instances on current neutral-atom platforms \cite{nguyen2023quantum,Bombieri2025,Oliveira2025,Schuetz2025}.

Second, although cliques in logical formulations require all-to-all connectivity among vertices associated with optimization variables, such connectivity need not be physically realized on a Rydberg platform. Our gadget enables the embedding of arbitrary incompatibility (MWIS) graphs into geometrically feasible atomic arrangements while preserving clique constraints. Large logical cliques can thus be encoded without long-range interactions or complex geometries, with the required connectivity enforced through blockade-mediated interactions within the gadget rather than direct physical coupling.

Third, the number of Rydberg atoms required to encode a problem with $m$ constraints and $N$ optimization variables scales as $\mathcal{O}(m N)$, which can represent a lower overhead than alternative constructions exhibiting $\mathcal{O}(N^2)$ scaling \cite{nguyen2023quantum}. The atom count can be further reduced by using an elimination and reordering procedure that leads to substantial hardware-resource savings, while preserving the encoded constraint structure.

Fourth, because the $xor_1$ gadget does not rely on problem-dependent detuning values or penalty terms, it eliminates the need for extensive classical preprocessing typically required for parameter tuning \cite{nguyen2023quantum}. The geometric reordering step constitutes only a lightweight compilation stage with negligible computational cost and therefore does not significantly contribute to preprocessing overhead.

%We demonstrate the applicability of the framework using representative optimization tasks, including airport gate assignment and the $N$-queens problem (Secs.~\ref{sec:gate} and \ref{sec:nqueens}). In both cases, we show that the ground state of the resulting Rydberg atom array encodes a valid solution to the corresponding optimization problems. We found that the maximum detuning range required using our gadget is \(6\) (see Table~\ref{tab:xorbenchmark} and Sec.~\ref{sec:scaling}), representing a reduction of approximately \(99\%\) relative to the detuning values required in the QUBO-based approach. In addition, we report on reductions of up to \(37\%\) in atom count compared to the QUBO-based approach for gate assignment problem and a \(54\%\) reduction for $N$-queens problem. 

We demonstrate the applicability of the framework using representative optimization tasks, including airport gate assignment and the $N$-queens problem (Secs.~\ref{sec:gate}, \ref{sec:gateadd}, and \ref{sec:nqueens}). In both cases, the ground state of the resulting RAA built using our formalism encodes a valid solution to the corresponding CSP. Across these illustrations, the maximum detuning range required by our gadget is 6, corresponding to an approximately 99\% reduction compared with QUBO-based encodings (see Table~\ref{tab:xorbenchmark}). We also observe reductions in atom count of up to 37\% for the gate-assignment instance and 54\% for the $N$-queens problem relative to the QUBO-based approach.
%(see Table~\ref{tab:queens_benchmark}).

The \(xor_1\) gadget provides a modular building block for encoding a broad class of constraint optimization problems whose structure can be expressed in terms of exactly-one or zero-or-one constraints. Any optimization problem that admits a formulation in which optimization variables are grouped into constraint sets enforcing exclusive or optional selection can, in principle, be decomposed into instances of the \(xor_1\) gadget and embedded into an RAA using the same clique-based construction. This modularity enables the systematic extension of the present methodology to a wide range of constrained optimization problems beyond the specific examples considered here.

From an operational standpoint, applying the proposed framework to an optimization problem proceeds as follows. 
One first identifies the binary optimization variables and enumerates the constraint sets that impose exactly-one or zero-or-one selection rules. 
Each such constraint set is then mapped onto an instance of the \(xor_1\) gadget, and the collection of gadgets is assembled into an RAA. 
The ground state of the resulting many-body Hamiltonian then encodes a valid solution to the original problem.

\section{Mathematical description of \texorpdfstring{$xor_1$}{xor1}} \label{sec:math}
\subsection{Graph Representation of Clique Problem}

\begin{figure}
    \centering
    \includegraphics[width=0.6\linewidth]{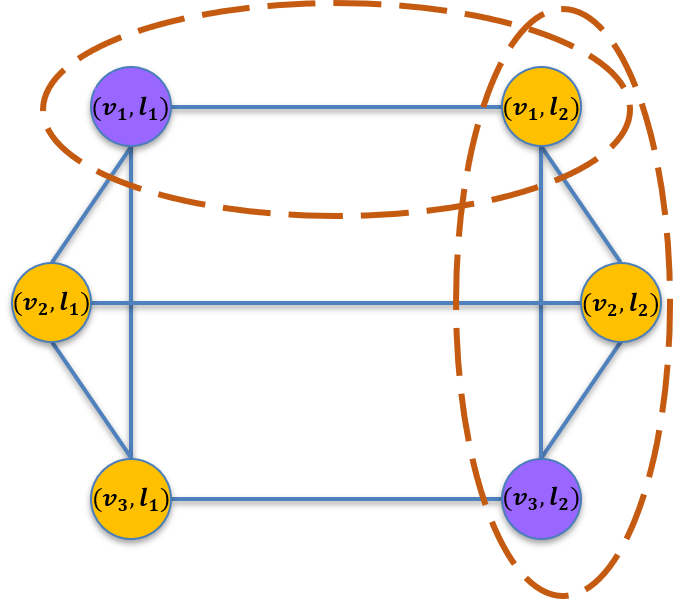}
    \caption{An example of cliques in an incompatibility graph for $\mathcal{V} = \{v_1, v_2,v_3\}$ and $\mathcal{L}=\{l_1,l_2\}$. Vertices correspond to variable-value assignments denoted by $(v_\alpha, l_\beta)$. Edges correspond to the disallowed mutual assignments. Cliques correspond to subsets of vertices that are fully connected, two of which are marked by dashed lines. The (un)selected assignments are marked by blue (orange) color corresponding to one of the feasible solutions.}
    \label{fig:clique}
\end{figure}

In our framework, graph-based representations, particularly incompatibility graphs, are used to encode CSPs and optimization problems with \textit{exactly-one} selection rules, such as airport gate assignment or $N$-queens problems. In these graphs, each decision variable $v_\alpha$ (with $v_\alpha \in \mathcal{V} = \{v_1, v_2, \dots\}$) is associated with a finite domain $\mathcal{L}=\{l_1,l_2,l_3,\dots\}$, and each vertex corresponds to a possible assignment of variable $v_\alpha$ to value $l_\beta$, denoted by $(v_\alpha, l_\beta)$.

We define an \emph{incompatibility graph} $\mathcal{G}_{\mathrm{incomp}}=(V,E)$, in which vertices represent variable-value assignments and edges connect mutually disallowed assignments. 
The definition of $\mathcal{G}_{\mathrm{incomp}}$ is given by
%A clique in $\mathcal{G}_{\mathrm{incomp}}$ corresponds to a set of mutually incompatible assignments 
\begin{equation}
\label{eq:incomp_graph}
\begin{split}
\mathcal{G}_{\mathrm{incomp}} &= (V,E),\\
V &= \{(v_\alpha,l_\beta)\mid v_\alpha\in \mathcal{V},\ l_\beta\in \mathcal{L}\},\\
E &= \bigl\{(v_\alpha,l_\beta)\leftrightarrow(v_\gamma,l_\zeta)\ \big|\ c_{\alpha \gamma}(l_\beta,l_\zeta)=\infty\bigr\},
\end{split}
\end{equation}
where $c_{\alpha \gamma}(l_\beta,l_\zeta)$ encodes the compatibility of jointly assigning $(v_\alpha,l_\beta)$ and $(v_\gamma,l_\zeta)$. In particular, $c_{\alpha \gamma}(l_\beta,l_\zeta)=\infty$ denotes a hard constraint violation. A clique in this setup corresponds to a subset of vertices that are fully connected in the incompatibility graph (see Fig. \ref{fig:clique}). 

In this representation, a feasible solution corresponds to an \emph{independent set} of vertices in $\mathcal{G}_{\mathrm{incomp}}$, i.e., a set of vertices no two of which are connected by an edge, ensuring that no incompatible assignments are selected simultaneously (see Fig. \ref{fig:clique}). To represent such selections algebraically, we introduce binary indicator variables $x_{\alpha,\beta} \in \{0,1\}$, where $x_{\alpha,\beta}=1$ denotes that the assignment $(v_\alpha,l_\beta)$, and hence the corresponding vertex $(v_\alpha,l_\beta)$ is selected. For instance, the \textit{exactly-one} constraint for each variable is enforced by requiring that exactly one assignment per variable is selected,
\begin{equation}
\sum_{\beta} x_{\alpha,\beta} = 1 \qquad \forall \alpha.
\end{equation}

The formulation explained above naturally maps to MWIS problems. As an example, the maximum common subgraph problem, which is known to be $NP$-complete
\cite{guidobene2024improved,hartmanis1982computers}, can be modeled within this
framework by assigning weights, $\delta_{\alpha,\beta}$, to vertices and solving
\begin{equation}
\begin{aligned}
\text{maximize}\quad & \sum_{\beta} \delta_{\alpha,\beta}\,x_{\alpha,\beta} \qquad \forall \alpha,\\
\text{subject to}\quad
& x_{\alpha,\beta}+x_{\gamma,\zeta}\le1 \quad
\forall (v_\alpha,l_\beta)\leftrightarrow(v_\gamma,l_\zeta)\in E,
\end{aligned}
\end{equation}
in which the objective maximizes the total weight of the selected vertices, while the constraints ensure that no two incompatible assignments, i.e., vertices connected by an edge in $\mathcal{G}_{\mathrm{incomp}}$, are selected. In what follows, we introduce a new method for solving constraint satisfaction problems using the clique-based formulation. For notational simplicity, we relabel the multi-index $(\alpha,\beta)$ as a single index $i$, writing $x_i \equiv x_{\alpha,\beta}$ and $\delta_i \equiv \delta_{\alpha,\beta}$, and refer to $x_i$ as optimization variables. In addition, configurations are represented as ordered tuples of binary variables, e.g., $(x_1,x_2,x_3,x_4)=(1,0,1,0)$ with each entry corresponding to the selection of a vertex.

%\shahram{Since our goal is to develop gadget-based constructions, we recast subsets of mutually exclusive assignments in terms of cliques of the incompatibility graph. A clique is a subset of vertices in which every pair is connected by an edge. In $\mathcal{G}_{\mathrm{incomp}}$, such a clique represents a set of assignments that are pairwise forbidden to occur simultaneously, as required for enforcing exactly-one or related exclusivity constraints. Any feasible configuration must therefore select at most one vertex from each clique, and admissible solutions correspond to independent sets that satisfy these clique-induced restrictions. This perspective forms the basis for the Boolean reformulation introduced in the following sections.}

\subsection{Reformulation of a Clique Problem as a Boolean Optimization: $xor_1$}

We reformulate clique-based constraints as a Boolean optimization problem using a Boolean function $xor_1$. This function enforces exactly-one selection over specified subsets of variables and provides a convenient algebraic representation for later mapping onto physical hardware.

%Let $\mathcal{X}$ be a set of optimization variables $x_i$, and let $\mathcal{Y} = \{\mathcal{Y}_1, \mathcal{Y}_2, \dots \}$ be a collection of non-empty subsets $\mathcal{Y}_k \subseteq \mathcal{X}$. Each subset is associated with a clique in the incompatibility graph, so that $\mathcal{Y}$ represents the total set of exactly-one constraints that must be satisfied.

Let $\mathcal{X}$ be the set of optimization variables $x_i$ and $\mathcal{Y}_k$ non-empty subsets of $\mathcal{X}$ associated with the cliques in the incompatibility graph. The collection $\mathcal{Y} = \{\mathcal{Y}_1, \mathcal{Y}_2, \dots \}$ of subsets $\mathcal{Y}_k$ represents all exactly-one constraints that must be satisfied.

For each subset $\mathcal{Y}_k$, we impose an exactly-one constraint through $xor_1$ which ensures that exactly one variable is active (corresponding to a binary value of 1), while all others remain inactive (binary value 0). This is written as
\begin{equation}
\label{eq:xor1}
xor_1(\mathcal{Y}_k) : \sum_{x_i \in \mathcal{Y}_k} x_i = 1,
\end{equation}
in which the sum is an ordinary integer sum over Boolean variables, not a sum modulo two. 
%Here, $xor_1(\mathcal{Y}_k)$ denotes the constraint, and the equality enforces that exactly one variable in the clique is 1. 
Throughout this work, the term $xor_1$ should not be confused with the Boolean XOR operation defined via modulo-two addition.
%This formulation is functionally equivalent to the \textit{exactly-one} selection constraint imposed by a clique in the compatibility graph. 
To express a zero-or-one constraint (i.e., allowing none or one active variable in $\mathcal{Y}_k$), 
we write ${xor}_1 (\{P\} \cup \mathcal{Y}_k)$ where $P$ is a dummy binary variable. It should be noted that each dummy variable is used only locally within a single constraint and does not interact with dummy variables associated with other subsets.
%we introduce a binary dummy set of variables $d_k=\{u_1,u_2,...\}$. Subsequently, Eq. \eqref{eq:xor1} is modified to
%\begin{equation}
%\label{eq:xor0}
   %{xor}_1 (d_k \cup \mathcal{Y}_k) : \sum_{x_i \in \mathcal{Y}_k, u_j \in d_k} x_i+u_j = 1 \implies 
   %\sum_{x_i \in \mathcal{Y}_k} x_i \leq 1,
%\end{equation}
%which ensures that either one variable in $\mathcal{Y}_k$ is active or none, with the activation of a dummy variable accounting for the "zero" case. Since each dummy variable is used only locally within a single constraint and does not interact with dummy variables associated with other subsets, its explicit index plays no role in the analysis. Therefore, we drop the explicit index and use $xor_1({d} \cup \mathcal{Y}_k)$ in which $d$ is an anonymous placeholder variable set used to control feasibility in the zero-or-one constraint.

\subsection{Multi-Constraint Reduction Using the \texorpdfstring{$xor_1$}{xor1} Encoding}

We now describe how the $xor_1$ formulation can be applied to reduce the complexity of problems involving multiple exactly-one constraints.
Let $\mathcal{Y}$ denote a collection of subsets, where each $\mathcal{Y}_k$ corresponds to a clique or constraint set in the underlying problem.
Applying the $xor_1$ operator to each subset yields the combined constraint
$\bigwedge_{\mathcal{Y}_k \in \mathcal{Y}} xor_1(\mathcal{Y}_k)$,
where $\bigwedge$ denotes the logical conjunction over all subsets.
This representation expresses multiple mutual-exclusion (or exactly-one) constraints in a unified Boolean form that can be directly incorporated into optimization or SAT-based formulations.
By exploiting shared variables or overlapping cliques, the $xor_1$ encoding can reduce both the number of explicit constraints and the total variable count, thereby lowering problem complexity without altering the feasible solution space:

%Given the complete set of binary variables $\mathcal{X} = \{x_1, x_2, \ldots\}$, each subset $\mathcal{Y}_j$ satisfies $\mathcal{Y}_j \subseteq \mathcal{X}$.
\textbf{I)} In the simplest case, if a subset $\mathcal{Y}_k$ contains only a single variable $x_i$, then all $xor_1$ involving $x_i$ can be resolved immediately:
the variable $x_i$ is fixed to~1, all other variables appearing with $x_i$ in the same constraints are set to~0, and these are subsequently removed from~$\mathcal{X}$.
After such a reduction, all parameter sets $\mathcal{Y}_j$ are updated to reflect the reduced variable set.
These reduction and synchronization steps are repeated until no further changes occur in either $\mathcal{X}$ or~$\mathcal{Y}$.
The next unresolved parameter set is then processed.

\begin{figure}
    \centering
    \includegraphics[width=0.45\linewidth]{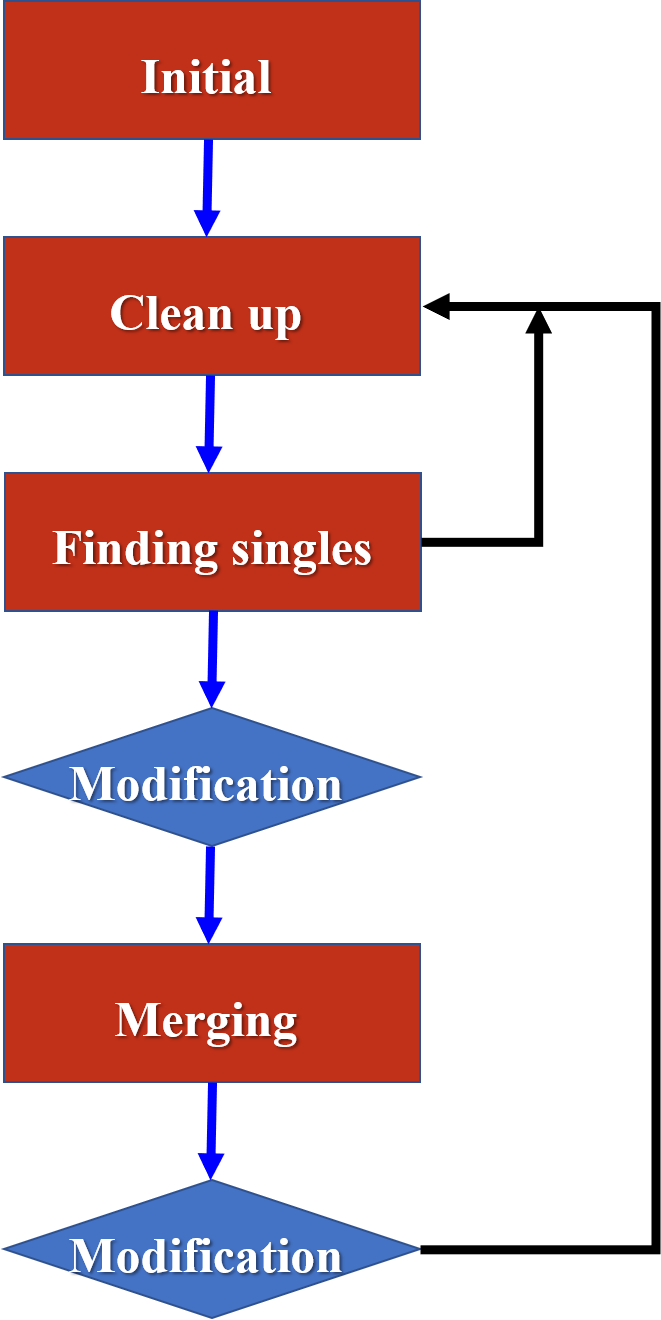}
    \caption{Processing multiple $xor_1$: $\bigwedge_{\mathcal{Y}_k \in \mathcal{Y}} xor_1(\mathcal{Y}_k)$. In the "Clean up", we remove zero-variables from parameter lists, then in "finding singles", we find the parameter lists with one value $x_{i}=1$ and their companions in other lists ($x_{o}=0$). After the modification in lists, parameter list inclusion takes place in which $\mathcal{Y}_1 \subsetneq  \mathcal{Y}_2 \Rightarrow \mathcal{Y}_2/\mathcal{Y}_1 :=0$.}
    \label{fig:multi_xor1_proc}
\end{figure}

\textbf{II)} If one subset, $\mathcal{Y}_1$, is fully contained within another, $\mathcal{Y}_1 \subsetneq \mathcal{Y}_2$, then satisfying $xor_1(\mathcal{Y}_1)$ automatically satisfies $xor_1(\mathcal{Y}_2)$
provided that all variables in $ \mathcal{Y}_2\setminus\mathcal{Y}_1$ are set to zero.
Those variables can therefore be removed from $\mathcal{X}$, and the parameter sets are again synchronized.
This process is iterated until no further reduction is possible (see Fig.~\ref{fig:multi_xor1_proc}).
Formally, the merge rule can be written as
\begin{equation}
\begin{aligned}  
& \mathcal{Y}_1 \subsetneq \mathcal{Y}_2, 
%\quad \mathcal{Y}_3 = \mathcal{Y}_2 \setminus \mathcal{Y}_1, 
  \\[2pt]
& xor_1(\mathcal{Y}_1) \land xor_1(\mathcal{Y}_2)
   \;\;\equiv\;\;
   xor_1(\mathcal{Y}_1) \land 
   \bigwedge_{x_i \in \mathcal{Y}_2 \setminus \mathcal{Y}_1} (x_i = 0),
\end{aligned}
\label{eq:xor1_merge}
\end{equation}
indicating that all variables $x_i \in \mathcal{Y}_2 \setminus \mathcal{Y}_1$ are fixed to zero and may be safely eliminated.

\textbf{Example.}
Consider the subsets
\begin{equation} \label{eq:example}
\mathcal{Y}_1 = \{x_1, x_2\}, \quad 
\mathcal{Y}_2 = \{x_1, x_2, x_3\}, \quad 
\mathcal{Y}_3 = \{x_3, x_4\}.
\end{equation}
Applying Eq.~\eqref{eq:xor1_merge} yields
\begin{equation}
\begin{aligned}
& xor_1(\{x_1, x_2\}) \land xor_1(\{x_1, x_2, x_3\}) \land xor_1(\{x_3, x_4\}) \\[2pt]
& \;\;\Rightarrow\;\;
   \big( xor_1(\{x_1, x_2\}) \land x_3 = 0 \big) \land xor_1(\{x_3, x_4\}) \\[2pt]
& \;\;\equiv\;\;
   xor_1(\{x_1, x_2\}) \land (x_3 = 0) \land (x_4 = 1).
\end{aligned}
\end{equation}
Hence, the feasible configurations are
\begin{equation}
(x_1, x_2, x_3, x_4)
  \in \{(0,1,0,1),~(1,0,0,1)\}.
\end{equation}
%Therefore, $x_3$ can be eliminated from the variable set, and $x_4$ is fixed to~1. 
This reduction procedure generalizes to any collection of overlapping $xor_1$s, systematically removing redundant variables while preserving the feasible solution space.

%The compact and local nature of the $xor_1$ formulation makes it particularly amenable to physical realization on Rydberg quantum computing platforms, where constraints can be directly encoded in the interaction graph of atom arrays. In the following section, we recast the problem within this framework and demonstrate how the $xor_1$ constraint can be implemented using Rydberg atom arrays and their native interaction dynamics. To this end, we first need to map cliques in our incompatatbility graphs to UDG and MWIS and then connect it to Rydberg platforms. 

The compact and local nature of the $xor_1$ formulation makes it well suited for physical realization on Rydberg platforms, where constraints can be directly encoded in the interaction graph of atom arrays. In the following section, we recast the problem within this framework and demonstrate how the $xor_1$ can be implemented as a gadget using native Rydberg interactions and atom arrangements. This requires mapping cliques in the incompatibility graph onto a UDG representation compatible with the MWIS formulation, and subsequently relating this representation to RAAs.

%\shahram{To be removed!}
%\textbf{Example.} 
%Consider the subsets
%\begin{equation}
%\mathcal{Y}_1 = \{x_1, x_2\}, \quad 
%\mathcal{Y}_2 = \{x_1, x_2, x_3\}, \quad 
%\mathcal{Y}_3 = \{x_3, x_4\}. \label{eq:example}
%\end{equation}
%Applying Eq.~\eqref{eq:xor1_merge} yields
%\begin{equation}
%\begin{aligned}
%& xor_1(\{x_1, x_2\}) \land xor_1(\{x_1, x_2, x_3\}) \land xor_1(\{x_3, x_4\}) \\[2pt]
%& \;\;\Leftrightarrow\;\; 
   %(xor_1(x_1, x_2) \land x_3 = 0) \land xor_1(x_3, x_4) \\[2pt]
%& \;\;\Leftrightarrow\;\; xor_1(x_1, x_2) \land (x_3 = 0) \land (x_4 = 1).
%\end{aligned}
%\end{equation}
%Hence, the feasible configurations are
%\begin{equation}
%\{(x_1, x_2, x_3, x_4)\}
 % = \{(0,1,0,1),~(1,0,0,1)\}.
%\end{equation}
%Therefore, $x_3$ can be eliminated from the variable set, and $x_4$ is fixed to~1.

%% merge(Xi, Yi)
%% merge-step-1: cleanup(Yi, Xi) -> Yi+1
%%               Yi+1 = copy(Yi)
%%               Yi+1 remove x not in Xi
%% merge-step-2: checksingles(Yi+1, Xi) -> Xi+1
%%               X+ = singles
%%               X- = partners
%%               Xi+1 = Xi/X-
%% merge-step-3: Yi != Yi+1  or Xi != Xi+1 -> i++ GOTO merge-step-1
%% merge-setp-4: merge(Yi+n, Xi+n) -> Yi+n+1, Xi+n+1
%%               Yi+n+1 = copy(Yi+n)
%%               yj in yk, x- = yk/yj -> remove yk
%%               X- = {x-} 
%%               Xi+n+1 = Xi+n/X-
%% merge-step-5: Yi+n != Yi+n+1 or Xi+n != Xi+n+1 -> GOTO merge-step-1

\section{Mapping Cliques to UDG and Rydberg Platforms} \label{sec:cliquetoRyd}

A UDG is a geometric graph that represents pairwise connectivity based on spatial proximity in the Euclidean plane.
Formally, let $V_{\mathrm{UDG}} = \{x_1, x_2, \ldots \}$ be a set of vertices corresponding to optimization variables, and associate with each vertex $x_i$ a position vector 
$\mathbf{r}_i \in \mathbb{R}^2$.
The corresponding UDG, $\mathcal{G}_{\mathrm{UDG}} = (V_{\mathrm{UDG}}, E_{\mathrm{UDG}})$ is defined by the edge set
\begin{equation}
E_{\mathrm{UDG}} =
\big\{(x_i,x_j)\,\big|\, 
\|\mathbf{r}_i - \mathbf{r}_j\| \le R
\big\},
\label{eq:udg_def}
\end{equation}
where $R>0$ is a fixed interaction (or connection) radius, which in the canonical case, $R=1$. To embed a clique from the incompatibility graph into a UDG, we place the vertices belonging to the clique within a region whose diameter does not exceed the interaction radius. This ensures that every pair of vertices lies within distance $R$, yielding the all-to-all connectivity required for a clique in the UDG. 
%Consequently, the edges induced within this region reproduce the incompatibility edges of the original graph, preserving the clique structure under the geometric embedding.
In this way, an individual clique from the incompatibility graph defined earlier can be represented locally in a geometric setting.
%This construction can be generalized to multiple cliques, which can be embedded as overlapping clusters in the UDG.

However, extending this construction to multiple cliques is nontrivial. In general, the embedding of several (possibly overlapping) cliques must satisfy global geometric consistency conditions imposed by the Euclidean metric. As a result, not every incompatibility graph admits a realization as a unit disk graph in two dimensions. In particular, independently specified clique constraints may induce conflicting distance requirements that cannot be simultaneously satisfied by any placement of points in \(\mathbb{R}^2\). Therefore, the mapping from a general incompatibility graph to a UDG is, in general, restricted. This is one of the issues that our gadget will address in the next section.

%Having established the UDG associated with a clique, we define the \emph{maximum independent set} problem on \(\mathcal{G}_{\mathrm{UDG}}\) as finding the largest subset of vertices \(\mathcal{S}\subseteq V\) such that no two vertices in \(\mathcal{S}\) are adjacent. The MWIS generalizes this formulation by assigning nonnegative weights \(\delta_i\) to vertices and seeking the independent set with maximum total weight. The MWIS Hamiltonian on a UDG can be written as \cite{nguyen2023quantum}
%\begin{equation}
%H_{\mathrm{MWIS}} = -\sum_i \delta_i x_i +  \sum_{(i,j)\in E_{\mathrm{UDG}}} U_{ij} x_i x_j,
%\label{eq:GeneralMWIS}
%\end{equation}
%where \(U_{i,j}\) is a Lagrange multiplier enforcing the independence constraint and, in our construction, the exactly-one constraint for each clique. The ground states of this Hamiltonian yield the MWIS solutions provided that \(U_{i,j} \gg \delta_i > 0\). These solutions select a consistent set of active vertices that maximizes the total weight while respecting all clique constraints.

We now briefly recall the MIS problem on a graph as finding the largest subset of vertices \(\mathcal{S}\subseteq V\) such that no two vertices in \(\mathcal{S}\) are adjacent. The MWIS generalizes this formulation by assigning nonnegative weights \(\delta_i\) to vertices and seeking the independent set with maximum total weight. In the context of UDGs (not necessarily restricted to those realizing multiple cliques), the MWIS problem can be written in Hamiltonian form as \cite{nguyen2023quantum}
\begin{equation}
H_{\mathrm{MWIS}} = -\sum_i \delta_i x_i + \sum_{(i,j)\in E_{\mathrm{UDG}}} U_{ij} x_i x_j,
\label{eq:GeneralMWIS}
\end{equation}
where \(U_{ij}\) is a Lagrange multiplier enforcing the independence constraint between adjacent vertices. The ground states of this Hamiltonian yield the MWIS solutions provided that \(U_{ij} \gg \delta_i > 0\). These solutions select a subset of vertices that maximizes the total weight while respecting all pairwise exclusion constraints encoded in the edge set.

%In Rydberg platforms, qubits are encoded in long-lived atomic states, with individual atoms controlled via optical tweezers and laser fields \cite{10.3389/frqst.2024.1426216,PRXQuantum.2.030322,Ebadi2021,browaeys2020,Scholl2021,pichler2018,Ebadi2022,Lanthaler2023,nguyen2023quantum}. Strong Rydberg interactions give rise to the blockade mechanism, which imposes a geometric constraint: two atoms separated by less than the blockade radius cannot be simultaneously excited. 

Rydberg-atom arrays provide a physical realization of such UDGs, in which the geometric arrangement of atoms directly determines the interaction graph. In these systems, qubits are encoded in long-lived atomic states, with individual atoms controlled via optical tweezers and laser fields \cite{10.3389/frqst.2024.1426216,PRXQuantum.2.030322,Ebadi2021,browaeys2020,Scholl2021,pichler2018,Ebadi2022,Lanthaler2023,nguyen2023quantum}. Strong Rydberg interactions give rise to the blockade mechanism, which imposes a geometric constraint: two atoms separated by less than the blockade radius cannot be simultaneously excited.

%This physical exclusion rule is mathematically equivalent to the clique constraint introduced earlier. When a set of atoms is placed inside one blockade region, the corresponding atoms lie within a unit disk, forming a clique in a UDG with atoms being the vertices in the UDG and edges in the UDG represented by physical blockade interactions between Rydberg atoms. Selecting an independent set of the UDG automatically implements the Boolean requirement that at most one element of each clique is active. Thus, any Boolean function built from the clique-based exactly-one condition can be encoded directly into the geometry of the atom array.

This physical exclusion rule is mathematically equivalent to the clique constraint introduced earlier. When a set of atoms lies within a blockade region, they form a clique in the corresponding UDG, with atoms as vertices and edges represented by blockade interactions. Selecting an independent set of the UDG thus automatically enforces the Boolean exactly-one constraint for each clique. 
%Consequently, any Boolean function constructed from clique-based exactly-one constraints can be directly encoded in the geometry of the atom array. 

Consider $N$ neutral atoms, each with a ground state $|g\rangle_i$ and a Rydberg (excited) state $|r \rangle_i$. In the rotating frame, the Hamiltonian under the rotating-wave approximation is 
\cite{pichler2018, Ebadi2022,Lanthaler2023,nguyen2023quantum}
\begin{equation}
\label{eq:rydberg_full}
H_{\mathrm{Rydberg}} = \frac{\Omega}{2}\sum_{i} \sigma_i^x - \sum_i \Delta_i\, n_i
    + \sum_{i<j} V_{ij}\, n_i n_j,
\end{equation}
where $\sigma_i^x$ coherently flips the state of atom $i$ from ground state to Rydberg state, $\Omega$ is the single-atom Rabi frequency, $n_i$ is the Rydberg-state occupation operator, $\Delta_i$ is the local laser detuning, and $V_{ij}$ is the van der Waals interaction between atoms $i$ and $j$, given by $V_{ij} = C_6 / \lVert \mathbf r_i - \mathbf r_j \rVert^6$, with $C_6$ the dispersion coefficient and $\mathbf r_i\in\mathbb{R}^2$ the atomic positions. In the classical limit, $\Omega \to 0$, the Hamiltonian~\eqref{eq:rydberg_full} maps directly onto the MWIS Hamiltonian in Eq.~\eqref{eq:GeneralMWIS} under the identifications $\delta_i = \Delta_i$ and $U_{ij} = V_{ij}$, with each vertex of the UDG corresponding to a Rydberg atom. Consequently, optimization on the device effectively corresponds to solving an MWIS problem on that UDG \cite{pichler2018, Ebadi2022,Lanthaler2023,nguyen2023quantum}. This is done by employing adiabatic protocols or QAOA-type variational algorithms, in which one steers the Rydberg array toward configurations that realize the MWIS, thereby solving MWIS instances defined by the underlying geometric layout \cite{nguyen2023quantum,Bombieri2025,Oliveira2025,Schuetz2025,Goswami_2024}.

The $xor_1$ formulation fits naturally into this framework: each clique can be realized as a tightly packed atomic cluster, with controlled geometric placement defining the required interactions between cliques. However, due to the geometric constraints discussed above, directly embedding an arbitrary incompatibility graph using only such local placements is, in general, not possible. 
%\sout{In this way, the $xor_1$ is directly mapped onto the physical UDG of the Rydberg array and enforced through blockade-mediated MWIS optimization. In practice, however, implementing this mapping is not entirely straightforward.} 
In practice, two additional limitations arise. First, only a finite number of atoms can be placed within a blockade radius of each other to form a clique. In a 2D array, this limits the size of each clique and prevents the straightforward implementation of large cliques solely through local placement. A second limitation concerns the implementation of weighted problems. In the Rydberg Hamiltonian, arbitrary positive vertex weights $\delta_i$ are encoded via local detunings, $\Delta_i = \delta_i$. Large dynamic ranges of $\Delta_i$ may be challenging experimentally, due both to the limited available detuning range and the requirement that $V_{ij} \gg \Delta_{\max}$ for adjacent sites.

%Our $xor_1$ gadget, which we introduce in the next section, circumvents these limitations. Through our design, we embed an arbitrary incompatibility/MWIS graph into a Rydberg geometry while preserving clique constraints, enabling arbitrary implementation of the cliques without requiring long-range interaction or complex higher-dimensional design. Second, all vertex weights, equivalently, all atomic detunings, remain as small as possible and within a narrow dynamic range. These required detunings do not change with the system size of the problem. These ensure that the gadget correctly encodes its logical structure while remaining compatible with feasible experimental detuning strengths with current state-of-the-art machines.

Our $xor_1$ gadget, introduced in the next section, circumvents these limitations. Its design allows us to embed an incompatibility graph into a Rydberg geometry while preserving clique constraints, enabling the implementation of large cliques without requiring all-to-all connectivity through long-range interactions or 3D arrangements. In addition, all vertex weights, equivalently, all atomic detunings, remain small and within a narrow dynamic range, which does not scale with the problem size. This ensures that the gadget correctly encodes its logical structure while remaining compatible with experimentally feasible detuning strengths on current state-of-the-art devices.

\begin{figure}[htbp]
    \centering
    \includegraphics[width=0.7\linewidth]{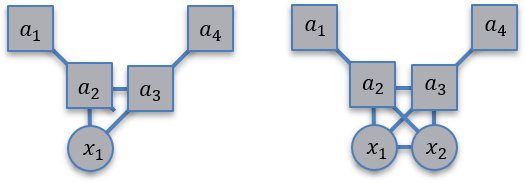} 
    \llap{\parbox[b]{5in}{(a)\\\rule{0ex}{0.8in}}}
    \includegraphics[width=0.98\linewidth]{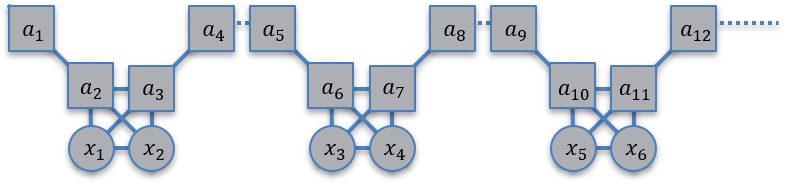}
    \llap{\parbox[b]{6.9in}{(b)\\\rule{0ex}{0.78in}}}
    \caption{(a) Unit cell of the $xor_1$ gadget for odd and even numbers of decision variables with gray color indicating that each vertex can be $\{0,1\}$, i.e. active or inactive. Here, $a_i$ denotes ancillary vertices, and $x_i$ is the optimization variable. Solid lines correspond to interactions within a unit cell (intracell interactions), while dotted lines denote interactions between different unit cells (intercell interactions). Both types of connections arise from Rydberg blockade and therefore impose hard constraints. A unit of the $xor_1$ gadget is constructed by taking the union of the graphs shown in the upper panel, or by having only one optimization vertex. Circles and squares represent vertices with weights $\delta$ and $2\delta$, respectively. (b) A multi-variable $xor_1$ gadget is built by taking connecting unit cells.
    }
    \label{fig:XOR}
\end{figure}

\begin{table}[h]
\centering
\small
\setlength{\tabcolsep}{4pt}
\begin{tabular}{|l|c|}
\hline
\textbf{Configuration} & \textbf{$W$} \\ \hline
\begin{tabular}[c]{@{}l@{}}
$x_i = 0$\\
$(a_1,a_2,a_3,a_4) \in \{(1,0,0,0),(0,1,0,0),(0,0,1,0),$\\
$(0,0,0,1)\}$ 
\end{tabular}
& $2\delta$ \\ \hline
\begin{tabular}[c]{@{}l@{}}
$x_i = 0$ \\
$(a_1,a_2,a_3,a_4) \in \{(1,0,0,1),(0,1,0,1),(1,0,1,0)\}$
\end{tabular}
& $4\delta$ \\ \hline
\begin{tabular}[c]{@{}l@{}}
$(x_1,x_2) \in \{(1,0),(0,1)\}$ \\
$a_i = 0$
\end{tabular}
& $\delta$ \\ \hline
\begin{tabular}[c]{@{}l@{}}
$(x_1,x_2) \in \{(1,0),(0,1)\}$ \\
$(a_1,a_2,a_3,a_4) \in \{(1,0,0,0),(0,0,0,1)\}$
\end{tabular}
& $3\delta$ \\ \hline
\begin{tabular}[c]{@{}l@{}}
$(x_1,x_2) \in \{(1,0),(0,1)\}$ \\
$(a_1,a_2,a_3,a_4) = (1,0,0,1)$
\end{tabular}
& $5\delta$ \\ \hline

\end{tabular}
\caption{All of the configurations with non-zero total weight for one unit cell of the $xor_1$ gadget.} \label{tab:conf}
\end{table}

\begin{figure}[htbp]
    \centering
    \includegraphics[width=0.98\linewidth]{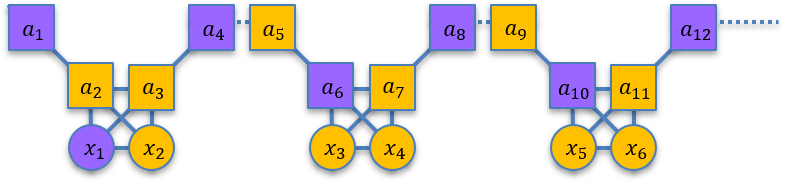}
    \llap{\parbox[b]{6.9in}{(a)\\\rule{0ex}{0.78in}}}   
    \includegraphics[width=0.98\linewidth]{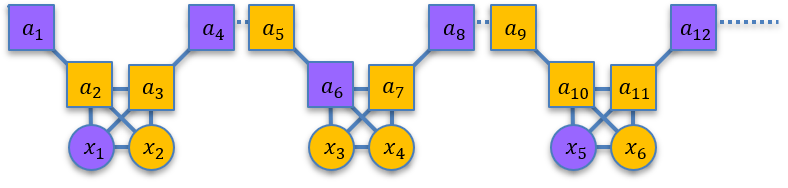}
    \llap{\parbox[b]{6.9in}{(b)\\\rule{0ex}{0.78in}}}    
    \caption{(a) A case in which only one optimization variable ($x_1$) is active (marked by blue color), and the rest of the optimization variables are inactive (marked by orange color). (b) a case in which two optimization variables are active. We observe that the upper panels admit higher weight compared to the lower panel, indicating that the upper panel yields one of the degenerate ground states of the Hamiltonian given in Eq. \eqref{eq:HamCostxor1}. }
    \label{fig:XORNew}
\end{figure}

\begin{figure}[htbp]
    \centering
    \includegraphics[width=1\linewidth]{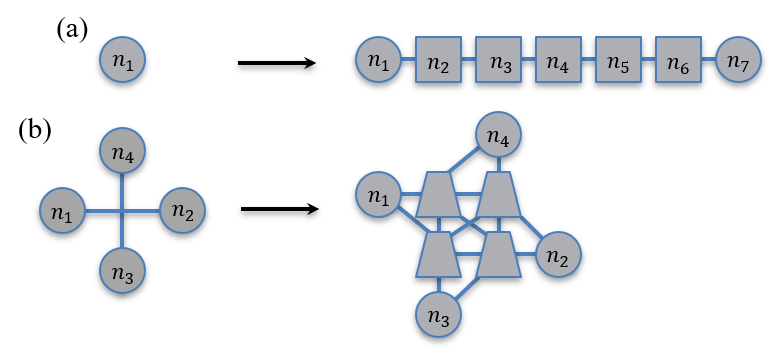} 
    \caption{(a) depiction of the copy gadget adapted from Ref.~\cite{nguyen2023quantum}. Odd vertices replicate the logical state of vertex $n_1$, enabling the propagation of its activity along a one-dimensional array. (b) depiction of the crossing gadget, which allows logical connections to pass through one another without inducing unwanted couplings. The trapezoids represent vertices with weights $4\delta$.}
    \label{fig:copyandcrossing}
\end{figure}

\begin{figure*}[htbp]
    \centering
    \includegraphics[width=1\linewidth]{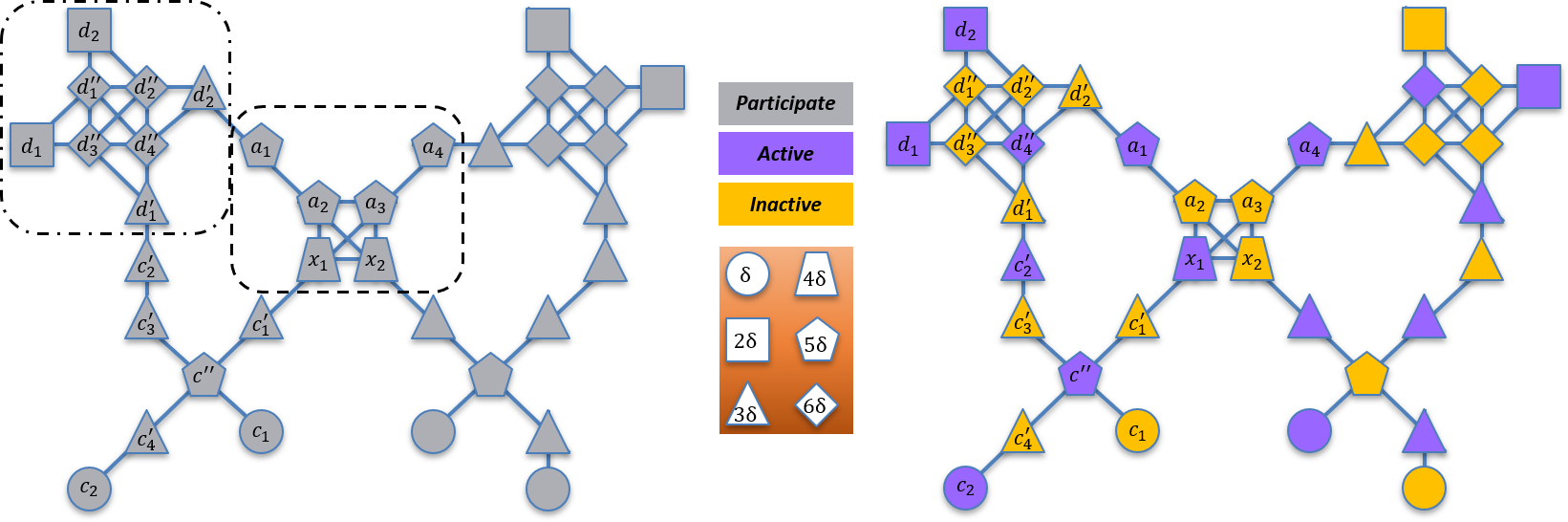}    
    \caption{Left panel: a composite structure of an $xor_1$ unit with required crossing and copy gadgets for two optimization variables. The circles, squares, triangles, trapezoids, pentagons, and diamonds correspond to having weight $\delta$, $2 \delta$, $3 \delta$, $4 \delta$, $5 \delta$, and $6 \delta$, respectively. The $d_i$ and $d_{i}^{\prime}$ are labeling exterior vertices of the crossing gadget while $d_{i}^{\prime\prime}$ the interior ones. Similarly, $c_2$, $c_i^{\prime}$, and $c_i^{\prime\prime}$ are labeling vertices of the copy gadgets. The unit of the $xor_1$ gadget with new weights is labeled by $x_i$ and $a_i$ and marked by a dashed black line (see Fig. \ref{fig:XOR} for more details), a black dotted-dashed line marks a crossing gadget, and the rest of the labeled vertices are part of copy gadgets except $c_1$. The vertices $c_{2}$, $c^{\prime}_{2}$, $c^{\prime\prime}$, and $d_{2}$ are different weighted copies of the $x_1$ vertex that encode its activity, which is evident in the right panel. The $d_1$ is a copy of $a_1$. The crossing gadget is applied for the crossing between $d_i^\prime$s and $d_i$s vertices. Right panel: the activation pattern that achieves the maximum weight of $54 \delta$ for one unit cell of the $xor_1$ gadget with one active optimization vertex. It should be noted that, to avoid overcrowding, we have partially presented the labels in the composite structures.}
    \label{fig:Cross-Copy}
\end{figure*}

\begin{figure}[htbp]
    \centering
    \includegraphics[width=1\linewidth]{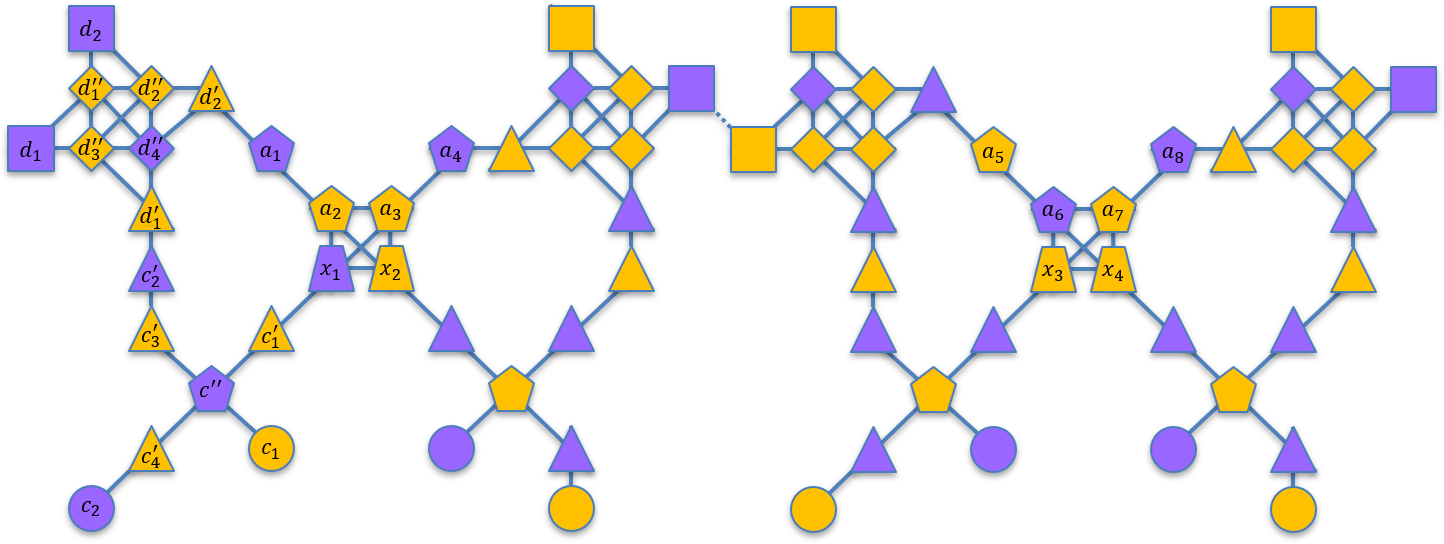}     
    \caption{Illustration of two composite structures containing two unit cells of the $xor_1$ gadget and hence four optimization vertices. We observe that the addition of crossing and copy gadgets with corresponding weights for each vertex faithfully realizes the exactly-one constraint. }
    \label{fig:Cross-Copy-general}
\end{figure}

\begin{figure}[htbp]
    \centering
        \includegraphics[width=1\linewidth]{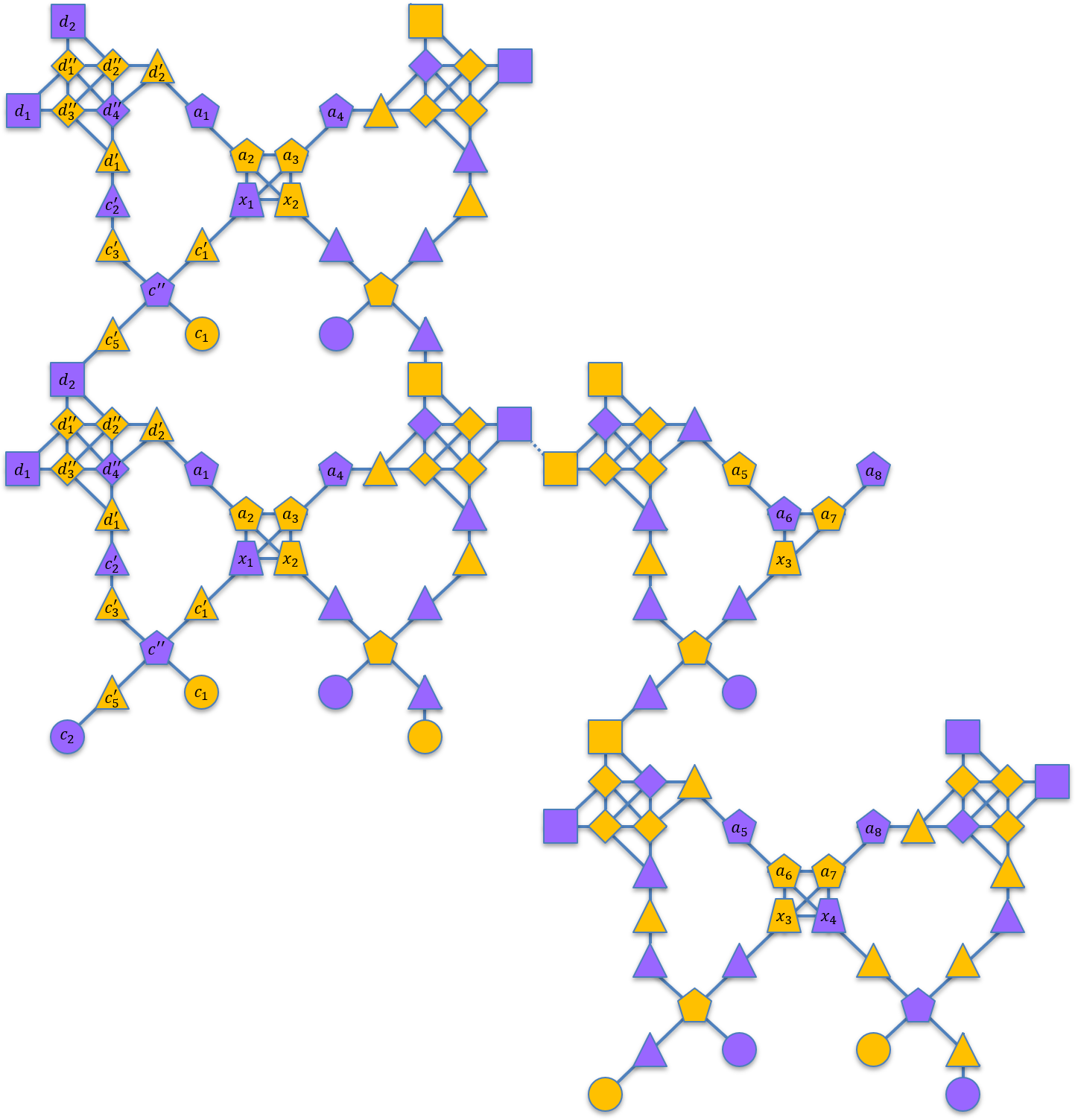}     
    \caption{The (non-optimized) graph realization of the example described in Eq. \eqref{eq:example} with composite structures. We observe that through the overlapping method, the activation of the one optimization vertex is transferred into other composite structures, realizing the conjunction between different $xor_1$ gadgets.}
    \label{fig:Cross-Copy-example}
\end{figure}

\begin{figure*}[htbp]
    \centering
    \includegraphics[width=0.75\linewidth]{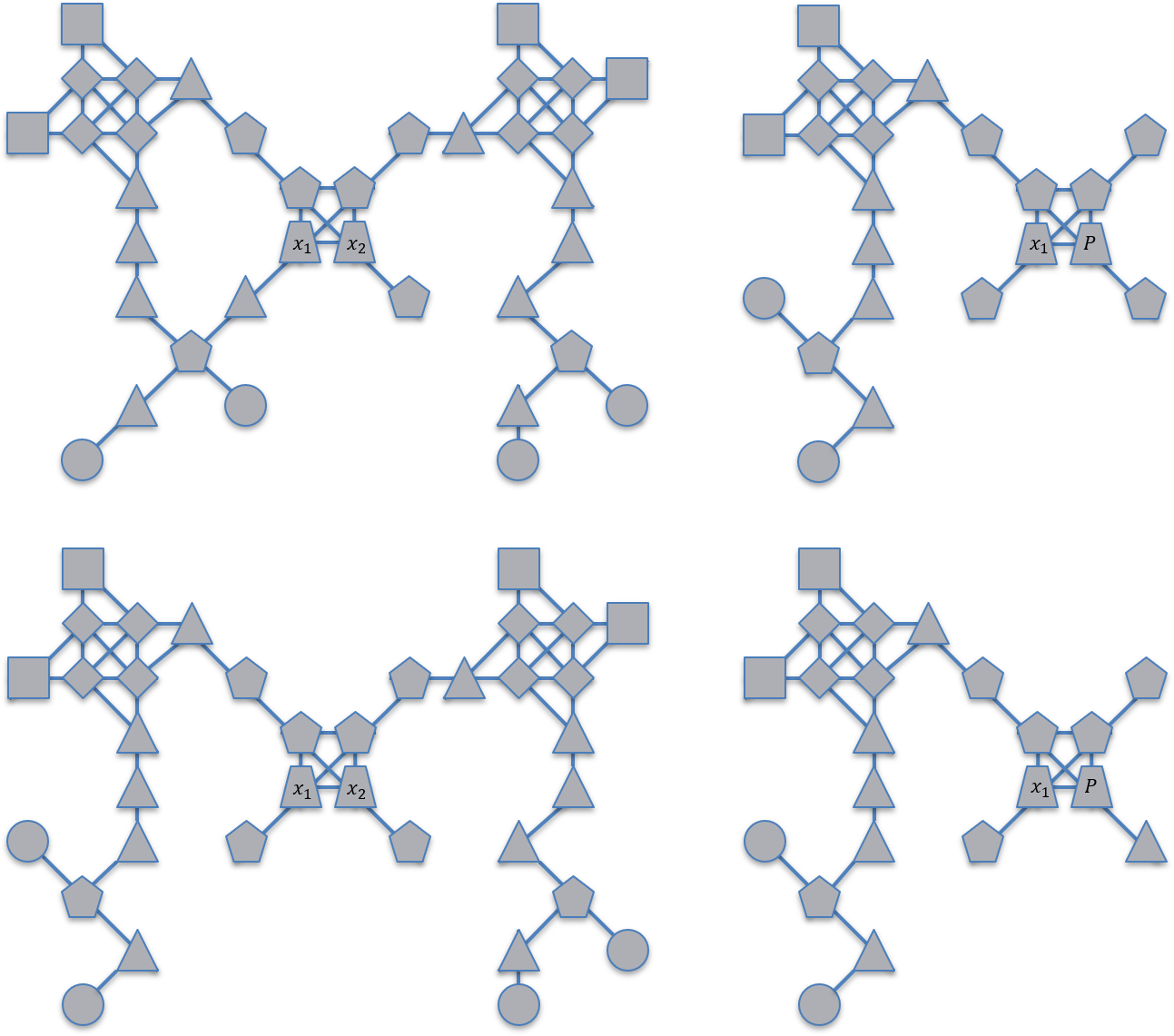} \llap{\parbox[b]{11in}{(a)\\\rule{0ex}{4.5in}}}   
    \llap{\parbox[b]{11in}{(b)\\\rule{0ex}{2.05in}}} 
    \llap{\parbox[b]{4.5in}{(c)\\\rule{0ex}{4.5in}}}   
    \llap{\parbox[b]{4.5in}{(d)\\\rule{0ex}{2.05in}}} 
    \caption{Additional composite structures used for the realization of the CSP in various scenarios. Panels (a)-(d) correspond to: (a) the $x_1$ participates in the CSP and $x_2$ does not, (denoted by $10$); (b) the optimization variables, $x_1$ and $x_2$, do not participate ($00$); (c) the optimization $x_1$ and the dummy variables do not participate ($0p$); and (d) the optimization variable $x_1$ does not participate while the dummy variable participates ($0P$). }
    \label{fig:additional}
\end{figure*}

\section{THE \texorpdfstring{$xor_1$}{xor1} GADGET AND COMPOSITE STRUCTURES} \label{sec:Xor1Alll}
\subsection{The structure of \texorpdfstring{$xor_1$}{xor1} gadget}  \label{sec:Xor1}

We now construct our $xor_1$ gadget within the UDG-MWIS framework to encode an exactly-one constraint over a set of optimization variables $\mathcal{Y}_k = \{x_1, x_2, \dots\}$. The structure of a unit cell of the $xor_1$ gadget is depicted in the Fig. \ref{fig:XOR}a. 
%Each unit cell contains one or two optimization vertices corresponding to one or two variables, $x_i$, with weight $\delta$ and four ancillary vertices, $a_i$, with weight $2\delta$, of which only two are connected with the optimization vertices. The ancillary vertices serve three purposes. 
Each unit cell contains one or two optimization vertices representing the optimization variables $x_i$, each with weight $\delta$, together with four ancillary vertices $a_i$ of weight $2\delta$, of which only two couple to the optimization vertices. It should be noted that the exactly-one constraint is imposed solely on the optimization vertices $x_i$, as will be shown by the fact that only one of these vertices becomes active in the $xor_1$ gadget. 
%First, they circumvent the limitation that an atom in a 2D grid can have at most eight direct neighbors. 
%First, they address the restriction of four vertices per clique without requiring long-range interactions, enabling embedding an arbitrary incompatibility or MWIS graph into a Rydberg geometry while preserving clique constraints (see details below) 

The ancillary vertices serve two purposes. First, they circumvent the geometric restriction on the number of vertices that can reside within a single blockade region. Through locally mediated constraints, they allow an arbitrary number of optimization vertices to be represented without requiring all-to-all physical connectivity necessary for the clique, while preserving the logical clique constraint under the embedding.
Second, the choice of a larger weight for ancillary vertices compared to optimization vertices will ensure that only one optimization vertex will be active in the gadget, hence ensuring the exactly-one constraint. For multiple variables $x_i$, we construct unit cells for each pair of variables and connect these cells via their ancillary vertices, as illustrated in Fig.~\ref{fig:XOR}b. As a result, for an $N$ optimization variables, we need $4\lceil \frac{N}{2} \rceil$ ancillary vertices. 

The Hamiltonian associated with this gadget is given by 
\begin{align}
   H_{\mathrm{xor}_1} &= - \delta \sum_{i=1}^N x_i - 2 \delta \sum_{j=1}^{4\lceil\frac{N}{2} \rceil} a_j+ U \bigg( \sum_{i=1}^{\lfloor N/2 \rfloor} x_{2i} x_{2i-1} + \notag 
   \\
   &
   \sum_{j=1}^{4\lceil\frac{N}{2} \rceil-1}  a_j a_{j+1} + \sum_{k=1}^{\lceil N/2 \rceil} (x_{2k} + x_{2k-1} )(a_{4k-1}+a_{4k-2}) \bigg), \label{eq:HamCostxor1} 
\end{align}
in which $U$ is the penalty energy, which is $U \gg 4 \delta >0$, ensuring that two neighboring vertices are not both active simultaneously. First, we need to show that the ground state of this gadget encodes the MWIS while admitting the exactly-one constraint, i.e., one optimization vertex is active. 

\textbf{\textit{Lemma:}} In the ground state, $xor_1$ gadget enforces exactly one active optimization vertex. \\
Let the gadget be composed of $N$ optimization vertices $x_1,\dots,x_N$ grouped into $M=\lceil N/2\rceil$ units (unit $k$ contains $x_{2k-1},x_{2k}$), and $4M$ ancilla vertices $a_1,\dots,a_{4M}$ arranged in unit blocks of four and linked as specified in the Hamiltonian of Eq.~\eqref{eq:HamCostxor1}. 
%Each optimization vertex carries weight $\delta$ and each ancilla carries weight $2\delta$. 
Then any MWIS (equivalently, any ground state of Eq.~\eqref{eq:HamCostxor1}) requires the activation of exactly one optimization vertex, and the ancillas follow the alternating pattern forced by that choice. The maximum total weight is
\begin{align}
W_{\max}=4M\delta+\delta,
\end{align}
and no independent set with $0$ or $\geq 2$ active optimization vertices reaches this value.

\textbf{\textit{Proof:}}\\
\textbf{(1) Restriction to independent sets.} 
Since $U \gg 4\delta$, any configuration containing adjacent active vertices incurs a penalty that dominates their weight contribution. 
Thus, the MWIS can be found by comparing independent sets only, with the maximum total weight given by
\begin{align}
W_{\max} = \delta\, m_x + 2\delta\, m_a,
\end{align}
where $m_x$ is the number of active optimization vertices and $m_a$ is the number of active ancillas. 

\textbf{(2) Local unit analysis.} 
Consider one unit with two optimization vertices and its four ancillas. 
Within a unit, two $x_i$s cannot be active simultaneously, and similarly, the maximum number of active ancillas among the four is $2$. 
%Considering that we need to maximize the weight and also satisfy the required constraint for the clique, the possible local contributions are:
%\begin{itemize}
    %\item No active optimization vertices, $x_i=0$ and two ancilla vertices are active ($(a_1,a_2,a_3,a_4) \in \{(1,0,0,1),(0,1,0,1), (1,0,1,0)\}$ for ancillary vertices), resulting in total weight $W=4\delta$. 
    %\item No active ancillary vertices, $a_i=0$, with only one optimization vertex being active ($(x_1,x_2) \in \{(1,0),(0,1)\}$), which leads to $W=\delta$.
    %\item One active ancillary vertex and one optimization vertex are active ($(a_1,a_2,a_3,a_4) \in \{(1,0,0,0),(0,0,0,1)\}$ and $(x_1,x_2) \in \{(1,0),(0,1)\}$), which gives us $W=3\delta$.    
    %\item Exactly one active optimization vertex $x_i$ ($(x_1,x_2) \in \{(1,0),(0,1)\}$) and two ancillary vertices ($(a_1,a_2,a_3,a_4)=(1,0,0,1)$ pattern for ancillary vertices) are active resulting in $W=5\delta$.
%\end{itemize}
All configurations with nonzero total weight are summarized in Table~\ref{tab:conf}. 
The maximum weight $5 \delta$ is obtained when the ancillas $a_1$ and $a_4$ and exactly one of $x_1$ or $x_2$ are active.

%In the following, we restrict our analysis to two cases: (i) configurations with one active optimization vertex and two active ancillary vertices, and (ii) configurations with no active optimization vertices and two active ancillary vertices. The remaining cases are omitted, as their generalizations to $N$ optimization vertices do not produce the maximum total weight. The second case is retained to explicitly demonstrate that its generalization likewise fails to attain the maximum total weight.

\textbf{(3) Global analysis.} 
In the absence of an active optimization vertex, the ancillas across all units form a chain, so their MWIS are exactly the two alternating patterns given by $(a_1,a_2,...) \in \{(0,1,0,1,0,...),(1,0,1,0,1,...)\}$
%with Hamiltonian cost function given by
%\begin{align}
   %H_{xor_1} &= - 2 \delta \sum_{j=1}^{4\lceil\frac{N}{2} \rceil} a_j, 
%\end{align}
and consequently, every unit contributes $4\delta$ resulting in total weight of
$W = 4M\delta$.

If exactly one optimization vertex is active in one of the units, the ancillary vertices in that unit are forced to take the activation form of $(1,0,0,1)$ (see Fig. \ref{fig:XORNew}a). This means that the unit contributes $5\delta$ to the total weight. This breaks the degeneracy between the two alternating patterns yielding a single overall pattern of ancilla vertices $(...0,1,0,1,0,0,1,0,1,...)$ that maximizes the weight to a total value of
%The ancillary pattern in this unit forces the other units to have the alternating ancillary pattern $(1,0,1,0)$ or $(0,1,0,1)$, depending on the placement of the unit and each contributing $4\delta$ to the total weight, resulting in 
\begin{align}
W = 5\delta + (M-1) 4\delta = 4M\delta + \delta, \label{eq:maxweight}
\end{align}
in which the first term on the RHS comes from the ancillary vertices and the second term is the contribution of the active optimization vertex. 
 
Activating a second optimization vertex elsewhere increases $m_x$ by $1$ (gain $\delta$) but forces at least two adjacent ancillas, in which at least one of them was active in the alternating pattern, to become inactive resulting in loss $= 2\delta$ with alternating pattern $(0,0,0,1)$ or $(1,0,0,0)$ for ancilla vertices of that unit. 
Thus, the net change is at most
\begin{align}
\Delta W = \delta - 2\delta = -\delta < 0,
\end{align}
indicating one unit less weight in that $xor_1$ unit (see Fig. \ref{fig:XORNew}b). The same can be repeated for an arbitrary number of active optimization vertices, and assuming that $N/2$ optimization vertices are active, we find the total weight as 
\begin{align}
W = 3 M\delta+ 2\delta,
\end{align}
which is smaller than what we found in Eq. \eqref{eq:maxweight}. This confirms that any configuration with $0$ or $\geq 2$ active optimization vertices has strictly smaller total weight than one with a single active $x_i$.
Therefore, the MWIS is uniquely realized (up to the symmetry associated with the choice of $x_i$ and the global ancillary parity pattern) by activating exactly one optimization vertex. The corresponding total weight is
$W_{\max}=4M\delta+\delta$, attained for $m_x=1$ and $m_a=2M$.

\subsection{Utilization of crossing and copy gadgets} \label{sec:crossing}

To employ $xor_1$ gadgets for multi-constraint satisfaction problems and physically realizing the conjunction ($\bigwedge_{\mathcal{Y}_k \in \mathcal{Y}} xor_1(\mathcal{Y}_k)$), we need to connect and synchronize $xor_1$ gadgets. We achieve this by introducing nontrivial modifications to the copy and crossing gadgets of Ref. \cite{nguyen2023quantum} and employing them in our construction as explained in the next paragraphs. The copy gadget replicates a logical variable across multiple physical sites while preserving coherence of the encoded state. Its structure is illustrated in Fig. \ref{fig:copyandcrossing}a, where each odd vertex encodes the state (e.g., activity) of the original vertex. The crossing gadgets are used to allow logical connections to pass through one another without inducing unwanted couplings (see Fig. \ref{fig:copyandcrossing}b).

The modifications we will introduce are guided by three criteria: (i) The exactly-one constraint enforced by the $xor_1$ gadget must be preserved. (ii) The copy and crossing gadgets must retain their intended functionality, while not perturbing the activation patterns of the $xor_1$ units. (iii) The weights for each vertex should be kept as small as possible, thereby allowing a larger portion of the energy spectrum to be allocated to encoding the cost function of the optimization problem.

Given the structure of the $xor_1$ gadget, achieving synchronization between distinct $xor_1$ gadgets requires, for each optimization vertex within a unit cell, one crossing gadget and two copy gadgets. Together, they form a composite structure shown in Fig.~\ref{fig:Cross-Copy}. In this structure, $c_{2}$, $c^{\prime}_{2}$, $c^{\prime\prime}$, and $d_{2}$ encode the activity of the corresponding optimization vertex, as they act as weighted copies of that vertex. The same applies to $d_1$, which encodes the activity of the $a_1$.

The first criterion requires that any composite structure containing a single active optimization vertex attains the maximal weight within the structure, while other structures with no active optimization vertex have exactly one unit lower total weight. To enforce this separation, the contribution of the $xor_1$ unit with an active optimization vertex must exceed that of any individual copy or crossing gadget within that structure. Accordingly, we increase the weights of the vertices in the $xor_1$ unit by three units. The weights in the auxiliary gadgets are then chosen to be sufficiently small so that they do not override the activation pattern of the $xor_1$ unit, yet large enough to partially compensate composite structures in which no optimization vertex is active, ensuring that the total weights of such structures remain exactly one unit lower. 

Concretely, the vertices in the copy gadgets ($c_i^{\prime}$) are assigned weights that are uniformly one unit smaller than those of the optimization vertices (see Fig. \ref{fig:Cross-Copy}). Two exceptions are required: (i) the intersection vertex of the copy gadgets (depicted by $c^{\prime \prime}$), which is assigned a larger weight due to overlap of copy gadgets, and (ii) the last vertex of one copy branch (depicted by $c_2$), which is similarly adjusted to preserve the intended propagation of activity. A similar weighting principle is applied to the crossing gadgets. Finally, the vertex labeled $c_1$ is introduced to guarantee that configurations without an active optimization vertex remain exactly one unit lower in total weight.

The total Hamiltonian associated with a composite structure is
\begin{align}
   H_{\mathrm{Total}} &= H_{\mathrm{xor}_1}+H_{\mathrm{Cross}}+H_{\mathrm{Copy}},
   \label{eq:TotHamCost} 
\end{align}
where $H_{\mathrm{xor}_1}$ is defined in Eq.~\eqref{eq:HamCostxor1} with the adjusted vertex weights, and $H_{\mathrm{Cross}}$ is the Hamiltonian of the crossing gadget given by 
\begin{align}
   H_{\mathrm{Cross}} &=\sum_i\bigg[ -2\delta \sum_{j=1}^2 d_{j}(x_i)-3\delta \sum_{j=1}^2 d^{\prime}_j(x_i)-  \notag 
   \\
   &
   6 \delta \sum_{j=1}^4 d^{\prime \prime}_j(x_i)+U\bigg(  d_{1}(x_i)d_{1}^{\prime\prime}(x_i)+d_{1}(x_i)d_{3}^{\prime\prime}(x_i)+ 
 \notag 
   \\
   &     
   d_{2}(x_i)d_{1}^{\prime\prime}(x_i)+d_{2}(x_i)d_{2}^{\prime\prime}(x_i)+d_{1}^{\prime}(x_i)d_{3}^{\prime\prime}(x_i)+ 
   \notag 
   \\
   & 
   d_{1}^{\prime}(x_i)d_{3}^{\prime\prime}(x_i)+ d_{2}^{\prime}(x_i)d_{2}^{\prime\prime}(x_i)+
   d_{2}^{\prime}(x_i)d_{4}^{\prime\prime}(x_i)+
   \notag 
   \\
   &    
    \sum_{j}d_i^{\prime\prime}(x_i)d_j^{\prime\prime}(x_i)\bigg]+d_{2}^{\prime}(x_1)a_1+d_{1}^{\prime}(x_1)c_{2}^{\prime}(x_1)+   
   \notag 
   \\
   &     
    d_{1}(x_2)a_4+d_{1}^{\prime}(x_2)c_{2}^{\prime}(x_2) \bigg), 
   \label{eq:Hamcopy} 
\end{align}
where $d_j(x_i)$ and $d^{\prime}_j(x_i)$ are exterior vertices (see Fig. \ref{fig:additional}) of the crossing gadget corresponding to optimization vertex $x_i$, with weights $2 \delta$ and $3 \delta$, respectively, and $d^{\prime \prime}_j(x_i)$ denote interior vertices of the same crossing gadget with weight $6\delta$. 
The last four terms originate from the connections between the crossing gadgets and the $xor_1$ unit, as well as from the copy gadgets. It should be noted that we need to adjust the value of $U \gg 12\delta$ to prevent violation of the blockade. Similarly, the Hamiltonian of the copy gadgets, $H_{\mathrm{Copy}}$ is given by
\begin{align}
   H_{\mathrm{Copy}} &= \sum_i\bigg[-\delta\sum_{j=1}^2c_{j}(x_i)-3 \delta \sum_{j=1}^4 c^{\prime}_{j}(x_i) -5 \delta c^{\prime \prime} (x_{i})+  
   \notag 
   \\
   &
   U\bigg(c_{1}(x_i)c^{\prime\prime}(x_i)+c_{4}^{\prime}(x_i)c_2(x_i)+c^{\prime\prime}(x_i)\sum_j c^{\prime}_{j}(x_i)-
   \notag 
   \\
   &       
   c^{\prime\prime}(x_i)c^{\prime}_{2}(x_i)+  
   c^{\prime}_{2}(x_i)c^{\prime}_{3}(x_i)+   
   c^{\prime}_{1}(x_i)x_i \bigg)\bigg], 
   \label{eq:Hamcoppy} 
\end{align}
where $c_{j}(x_i)$, $c^{\prime}_{j}(x_i)$, and $c^{\prime\prime}_{j}(x_i)$ denote vertices of the copy gadget associated with $x_i$, carrying weights $\delta$, $3\delta$, and $5\delta$, respectively (see Fig.~\ref{fig:Cross-Copy}). 

In the next chapter, we verify that these weight assignments indeed enforce the exactly-one constraint. By analyzing MWIS for individual and connected composite structures, we show that any configuration violating this rule has strictly lower total weight. This confirms that the addition of copy and crossing gadgets does not alter the logical function of the $xor_1$ units and ensures global consistency across multiple connected gadgets.

\subsection{MWIS for the composite structure}

%Determining the MWIS of the full composite structure reduces to maximizing the total weight within each gadget while enforcing consistency across their interconnections. 
The maximum weight attainable by a single $xor_1$ unit with new weight assignments is $14\delta$. 
For each crossing gadget, the maximum weight lies between $10\delta$ and $12\delta$, depending on the activity of the associated optimization vertices in the $xor_1$ unit. 
Similarly, the copy gadgets corresponding to each optimization vertex attain maximum weights of $9\delta$ or $10\delta$, again depending on the optimization vertex being active or inactive.

In the case of one active optimization vertex in the $xor_1$ unit, the activation pattern of vertices within this unit cell directly constrains the activity of vertices in the copy and crossing gadgets to preserve the global maximum-weight configuration. The maximum total weight for the composite structure, shown in Fig. \ref{fig:Cross-Copy}, with one active optimization vertex is $54 \delta$. This value is obtained as follows: the copy gadgets associated with the active and inactive optimization vertices contribute $9\delta$ and $10\delta$, respectively, while the corresponding crossing gadgets contribute $10\delta$ and $11\delta$.

For multiple optimization vertices, we construct composite structures from $xor_1$ unit cells using the same procedure and connect them through their crossing gadgets, as illustrated in Fig.~\ref{fig:Cross-Copy-general}. Activation of an optimization vertex within a unit of $xor_1$ and subsequently its composite structure dictates the admissible configurations in neighboring crossing gadgets, propagating constraints across adjacent units. This is shown below for two composite structures, which can be generalized to an arbitrary number of composite structures.

Consider two composite structures involving four optimization vertices, and suppose that one of the structures contains a single active optimization vertex. The resulting activation pattern within that structure is then fixed as described previously. 
At the interface between the two structures, where their crossing gadgets intersect, the activity of vertices in the crossing gadget of the structure with active optimization vertex restricts the possible configurations of the adjacent crossing gadget belonging to the neighboring structure. In particular, only one configuration yields the maximum weight in that crossing gadget: 

One interior vertex and two exterior vertices with weights $3\delta$ are active (see Fig. \ref{fig:Cross-Copy-general}). 
In this configuration, the $\mathit{xor}_1$ unit contributes a weight of $10\delta$, with no optimization vertices active, while the copy gadgets together with the remaining crossing gadget contribute $41\delta$, yielding a total weight of $53\delta$. Any other configuration, in which one optimization vertex is active, yields at maximum a total weight of $52\delta$, one unit smaller than the optimal configuration. Moreover, the activation pattern of the rightmost crossing gadget coincides with that of the composite structure in which the first unit has an active optimization vertex. Consequently, the same activation pattern is propagated to the next composite structure. This confirms that the MWIS is obtained when the exactly-one constraint is satisfied and demonstrates that the addition of the copy and crossing gadgets does not alter the logical function of $xor_1$ units. By extending this construction to an arbitrary number of composite structures, the same rule holds.

\subsection{The realization of the conjunction in composite structures}

%To construct logical conjunctions between multiple $\mathit{xor}_1$ gadgets, we introduced the combined use of the copy and crossing gadgets.  
%These auxiliary structures ensure that the activation pattern of one $\mathit{xor}_1$ unit propagates correctly to the others while preserving the exactly-one constraint.  
The conjunction between $\mathit{xor}_1$ structures can be implemented via the overlapping method.
In this method, the legs of the copy gadgets associated with each optimization vertex are \emph{overlapped} with the exterior vertex corresponding to the same optimization variable in the crossing gadget of the subsequent $\mathit{xor}_1$ structure.  
Each pair of overlapping vertices is assigned an effective weight of $2\delta$ (see Fig. \ref{fig:Cross-Copy-example} for more details).  
%\textbf{The MWIS is therefore obtained by simultaneously maximizing (i) the number of active overlapping vertices across the entire structure and (ii) the total weight within each $\mathit{xor}_1$ composite structure (including crossing and copy gadgets).}

The activation pattern in one composite structure naturally enforces a consistent configuration in adjacent composite structures as well as other composite structure units belonging to other $xor_1$ structures through the overlapping vertices, thereby maintaining the global exactly-one constraint. 
%This is expected as our modified crossing and copy gadgets are designed to coherently transfer or replicate activation patterns. 
To demonstrate this, we consider the example defined in Eq.~\eqref{eq:example}, whose structural realization is shown in Fig.~\ref{fig:Cross-Copy-example}. 
The two optimal solutions yield a total
weight of $192 \delta$, with $(x_1, x_2, x_3, x_4) = (1,0,0,1)$ shown in Fig.~\ref{fig:Cross-Copy-example}.  
All other configurations, which are illegal, lead to smaller total weights, confirming that this assignment globally maximizes the combined weight of all $\mathit{xor}_1$ structures. For example, if $x_3$ becomes active in Fig.~\ref{fig:Cross-Copy-example} and one keeps one of the $x_1$ or $x_2$ active, the maximum total weight will be $190 \delta$, which is two units smaller compared to the correct constraint-satisfying degenerate solution. 

%A practical advantage of the overlapping method is that, in the context of Rydberg-atom implementations, measurements are required only on the overlapping atoms to determine the logical output of the computation. The rest of the array does not need to be scanned, which significantly simplifies experimental readout.

%\noindent\textbf{II) Stacking method.}  
%Alternatively, the conjunction between consecutive $\mathit{xor}_1$ structures can be established by stacking the $xor_1$ structures on top of each other instead of direct overlaps. This requires specific modifications in the weights of the vertices.  
%As in the overlapping case, the MWIS is obtained by maximizing the total weight within each $\mathit{xor}_1$ unit independently. 
%We will not focus on this method in this paper, as the results will be the same with additional vertices and modifications to the weight in each vertex. 

\subsection{Scaling analysis} \label{sec:scaling}

%In general, for each optimization vertex, as was stated before, we need $4$ ancillary vertices, which are shared by another optimization vertex in the unit cell of the $xor_1$ gadget, resulting in $4\lceil \frac{N}{2} \rceil$ vertices for $N$ optimization variables. 

For $N$ optimization variables, we have $4\lceil \frac{N}{2} \rceil$ vertices due to $xor_1$ gadget.
Each crossing gadget contains $8$ vertices, and for the copy gadgets, we need $7$ vertices. Therefore, in total for $N$ optimization vertices, we need $N_{total}=N+4\lceil \frac{N}{2} \rceil+8 N+7N=16N+4\lceil \frac{N}{2} \rceil$. Subsequently, for $m$ constraints and considering the overlapping, we need $m \times N_{total}-(m-1)N$ vertices (Rydberg atoms). It should be noted that in practice, it is possible to reduce the number of atoms by using the algorithms given in Fig. \ref{fig:multi_xor1_proc} and the reordering method, which we will use in the illustration example. The highest required number of connections between the vertices is $4$ within the crossing and $xor_1$ gadgets, and the highest detuning in a vertex is $6\delta$ (for interior vertices of the crossing gadget), putting our $xor_1$ structure within the reach of the current state-of-the-art Rydberg quantum computing platforms.

\subsection{From Constraint Satisfaction to Optimization}

%For the Hamiltonian $H_{\mathrm{Total}}$, the ground-state manifold is degenerate, corresponding to all feasible configurations satisfying the encoded constraints. Optimization is naturally incorporated by introducing a cost function Hamiltonian, $H_{\mathrm{cost}} = f(x_1, x_2, \dots, x_N)$ and forming the Hamiltonian
%\begin{equation}
%H = H_{\mathrm{Total}} + H_{\mathrm{cost}}.
%\end{equation}
%Here, $H_{\mathrm{cost}}$ assigns weights to the optimization vertices in the MWIS graph, effectively biasing the system toward the optimal feasible solution. For example, a linear objective can be implemented as $H_{\mathrm{cost}} = -\sum_i \delta^\prime_i x_i$, producing a re-weighted MWIS problem whose ground state maximizes the total weight among all independent sets that satisfy the gadgets' constraints.

%However, in contrast to the constraint Hamiltonian, the total variation of $H_{\mathrm{cost}}$ over all configurations typically increases with system size. As a result, directly adding $H_{\mathrm{cost}}$ can distort the relative energy structure imposed by $H_{\mathrm{Total}}$, even when blockade constraints remain satisfied. To preserve the logical structure of the constraints under the addition of $H_{\mathrm{cost}}$, the energy penalty for violating any constraint must dominate the variation of the cost function over feasible configurations.

For the Hamiltonian $H_{\mathrm{Total}}$, the ground-state manifold is degenerate, corresponding to all feasible configurations satisfying the encoded constraints. Optimization can be incorporated by introducing a cost function Hamiltonian, $H_{\mathrm{cost}} = f(x_1, x_2, \dots, x_N)$. However, if the total variation of $H_{\mathrm{cost}}$ becomes too large, directly adding $H_{\mathrm{cost}}$ can distort the relative energy structure imposed by $H_{\mathrm{Total}}$, even when blockade constraints remain satisfied. To preserve the logical structure of the constraints under the addition of $H_{\mathrm{cost}}$, the energy penalty for violating any constraint must dominate the variation of the cost function over all feasible configurations.

We formalize this requirement in terms of the constraint gap. Let $\zeta$ denote the minimum energy penalty associated with violating the logical constraints, i.e., the spectral gap separating the feasible manifold from the lowest-energy constraint-violating configurations. In the present construction, this gap arises from the interplay between the interaction scale $U$ and the detuning scale $\delta$. Each violation of a blockade constraint incurs an energy penalty of at least $U$, while detuning contributions modify energies by at most $\mathcal{O}(\delta)$ per site. Consequently, for $U \gg 12\delta$, the constraint gap satisfies $\zeta \ge U - 12 \delta$. To preserve the constraint manifold under the addition of the cost function, one must ensure that this gap dominates the variation of the objective function within the feasible solution manifold $\zeta \gg \max_x f(x) - \min_x f(x)$. In practice, this condition can be enforced in two conceptually equivalent ways:

\begin{enumerate}
\item \textbf{Cost function rescaling:} One introduces a scaled cost Hamiltonian
\begin{equation}
H = H_{\mathrm{Total}} + \epsilon H_{\mathrm{cost}},
\end{equation}
with $\epsilon$ chosen such that the variation of $\epsilon H_{\mathrm{cost}}$ remains small compared to $\zeta$. This ensures feasibility is preserved while optimization acts as a perturbation on the constrained subspace.
\item \textbf{Global constraint rescaling:} Alternatively, one may rescale all parameters of the constraint Hamiltonian, including both detuning terms and interaction strengths, according to
\begin{equation}
H = \lambda H_{\mathrm{Total}} + H_{\mathrm{cost}},
\end{equation}
where $\lambda$ is chosen such that the minimum energy cost of any constraint violation exceeds the maximal variation of $H_{\mathrm{cost}}$. This rescaling preserves the relative structure of the xor$_1$, copy, and crossing gadgets, while maintaining the blockade condition.
\end{enumerate}

As an alternative to explicit cost-function encoding, one may incorporate optimization by introducing additional auxiliary structures associated with each optimization variable. In this approach, the cost function effectively reweights the optimization vertices in the MWIS graph, as in $H_{\mathrm{cost}} = -\sum_i W_i x_i$. An optimization variable $x_i$ with weight $W_i$ is then assigned $W_i$ identical auxiliary contributions, each providing a fixed energy bias when $x_i = 1$. This construction reproduces the effect of the cost function, favoring configurations that maximize the total number of activated contributions. While conceptually equivalent to introducing $H_{\mathrm{cost}}$, this approach requires a number of additional degrees of freedom that scales with the magnitude of the weights, leading to increased resource overhead. As a result, direct detuning-based encoding of the cost function may provide a more efficient and scalable implementation for the current state-of-the-art machines.

With these approaches, the Hamiltonian retains a clear hierarchy of energy scales: constraint violations are suppressed by the largest scale (set by $U$), the logical structure of the gadgets is enforced at an intermediate scale (set by $\delta$), and the cost function acts as a weak perturbation selecting the optimal configuration within the feasible manifold.

In this way, the MWIS structure of $H_{\mathrm{Total}}$ is preserved while enabling optimization. Compared to penalty-based constructions, which often require constraint strengths that grow with problem size, the present approach benefits from fixed, problem-size-independent detuning and interaction scales. This allows the available energy range to be used more efficiently, while still requiring only a controlled global rescaling to incorporate the cost function. Consequently, the $xor_1$-based construction provides a scalable framework in which constraint satisfaction is enforced through Rydberg blockade and gadget design, and optimization is implemented via a tunable cost term acting within the constrained manifold.

\section{Application of \texorpdfstring{$xor_1$}{xor1} gadget for Gate assignment problem} \label{sec:gate}
A representative application of our proposed gadget is the \emph{airport gate assignment problem}, in which incoming aircraft are assigned to available gates. Two classes of restrictions are fundamental. First, \emph{compatibility restrictions} arise from physical or operational requirements, such as aircraft size or passenger-handling needs (e.g., passport control), which may render certain flight-gate assignments infeasible. Second, \emph{temporal restriction} arises from overlapping arrival and departure time windows, which prevent multiple flights from being served simultaneously at the same gate.

We introduce binary variables \(x_{f,g}\) indicating the assignment of flights to gates,
\[
x_{f,g} =
\begin{cases}
1, & \text{if flight } f \text{ is assigned to gate } g, \\
0, & \text{otherwise}.
\end{cases}
\]

The instance is specified by the following sets:
\begin{itemize}
    \item \textbf{Flights:} \(F=\{f_1,f_2,\dots\}\), the set of arriving aircraft.
    \item \textbf{Gates:} \(G=\{g_1,g_2,\dots\}\), the set of available gates.
\end{itemize}

From these, we define two collections encoding scheduling restrictions:
\begin{itemize}
    \item \textbf{Slot conflicts:} In which we group flights that cannot occupy the same gate simultaneously  
    \[
    Q_1=\big\{Q_{1,i}\subseteq F \,\big|\, \text{flights in } Q_{1,i} 
    \text{ have overlapping time slots} \big\},
    \]

    \item \textbf{Incompatibilities:} In which we group the forbidden flight–gate assignments  
    \[
    Q_2=\big\{(f_i,g_j)\,\big|\, \text{flight } f_i 
    \text{ cannot be assigned to gate } g_j \big\},
    \]
\end{itemize}

Using the slot conflicts restriction, for every $f\in F$, we have
$\sum_{g\in G} x_{f,g}=1 ,$
which can be expressed using $xor_1$ given in Eq.~\eqref{eq:xor1} as
\begin{equation}
xor_1\big(\{x_{f,g} \mid g \in G\}\big):  \sum_{g \in G} x_{f,g} = 1,
\qquad \forall f \in F.
\end{equation}

Furthermore, for each gate, at most one flight from any mutually overlapping set may be assigned to that gate. This zero-or-one constrain is imposed for all $g\in G$ and $Q_{1,i}\in Q_1$ through an equality formulation with dummy variable $P$,
\begin{equation}
\begin{split}
&xor_1\big( \{ P \} \cup \{ x_{f,g} \mid f \in Q_{1,i} \} \big): \sum x_{f,g}+P  = 1 , \\
&\forall Q_{1,i} \in Q_1,\; \forall g \in G .
\end{split}
\end{equation}
where $d$ enforce the at-most-one assignment among conflicting flights. Finally, flight–gate incompatibilities are handled in a preprocessing step by fixing
\begin{equation}
x_{f,g} \coloneqq 0 ,
\qquad \forall (f,g) \in Q_2 .
\end{equation}
which is equivalent to removing these variables from the optimization model.

The presence or absence of dummy variables, along with the fact that an optimization variable may appear in some constraints but not others, necessitates the introduction of additional gadgets. To maintain synchronization and enforce the correct conjunction across all constraints, we introduce four additional gadgets shown in Fig.~\ref{fig:additional}. These gadgets correspond to the following cases:
%
%\begin{itemize}
  %\item $10$ and $01$: Corresponding to the case in which only one of the optimization vertices in the $xor_1$ gadget participates in the CSP while the other does not (see Fig. \ref{fig:additional}a),
  %\item $00$: Indicating neither optimization vertex in the $xor_1$ gadget participates (see Fig. \ref{fig:additional}b),
  %\item $0p$: Neither dummy nor optimization vertices participate (see Fig. \ref{fig:additional}c).
  %\item $0P$: Only a dummy vertices participates (see Fig. \ref{fig:additional}d),  
%\end{itemize}

\begin{itemize}
  \item $10$ and $01$: For the pair of optimization vertices in the unit cell of the $xor_1$ gadget, exactly one participates in the CSP ( Fig.~\ref{fig:additional}a).
  \item $00$: No optimization vertex in the unit cell participates in the CSP (Fig.~\ref{fig:additional}b).
  \item $0p$: No optimization or dummy vertex in the unit cell participates in the CSP (Fig.~\ref{fig:additional}c).
  \item $0P$: Only the dummy vertex in the unit cell participates in the CSP (Fig.~\ref{fig:additional}d).
\end{itemize}
Here, dummy variables, when present, are denoted by a trailing uppercase \texttt{P} (and lowercase \texttt{p} otherwise). The $11$ case, in which two optimization vertices participate, was defined previously (see Fig.~\ref{fig:Cross-Copy}). These constructions are not restricted to the gate-assignment problem and extend to a broader class of CSPs. Their validity follows directly from the arguments used for the $11$ case. The locality of the dummy variable is evident in their gadgets, as they contain neither crossing nor copy gadgets (see Fig.~\ref{fig:additional}d), and its unit cell always sits at the end of the $xor_1$ gadget, which is shown in Fig. \ref{fig:RAAA_minimal_4}.

By arranging these templates appropriately, we obtain the composite multi-$xor_1$ structure, which encodes all constraints in a geometry compatible with UDG-MWIS embedding. To illustrate this, we give a minimal example in the next section. 

\subsection{Minimal Example: four flights and three gates}

We consider a minimal illustrative example with three gates, \(G = \{A, B, C\}\). Gate~A can accommodate wide-body aircraft, whereas gates~B and~C are restricted to narrow-body aircraft. The required ground time is three time slots for wide-body aircraft and two time slots for narrow-body aircraft.

Four flights are scheduled. Flight~1 is operated by a wide-body aircraft and arrives at time slot~0. Flight~2 is operated by a narrow-body aircraft and also arrives at time slot~0. Flight~3 arrives at time slot~1, and flight~4 arrives at time slot~2. In the initial stage, for each flight-gate combination, we need to create one optimization variable, which results in
\begin{align}
\mathcal{X} = \{x_{1,A}, x_{1,B}, x_{1,C}, x_{2,A}, x_{2,B}, x_{2,C}, 
\notag \\
x_{3,A}, x_{3,B}, x_{3,C}, x_{4,A}, x_{4,B}, x_{4,C}\} .
\end{align}
Due to flight-gate incompatibilities, the variables \(x_{1B}\) and \(x_{1C}\) are infeasible and can be ignored. The remaining optimization variables are therefore
\begin{align}
\mathcal{X} &= \{x_{1,A}, x_{2,A}, x_{2,B}, x_{2,C},
\notag \\
& x_{3,A}, x_{3,B}, x_{3,C}, x_{4,A}, x_{4,B}, x_{4,C}\} .
\end{align}

%In the binary-assignment representation, variables that do not participate in an $xor_1$ constraint are denoted by a 0. 
Based on the arrival times and ground times, the time-slot–conflicting flight groups are \(\{f_1,f_2,f_3\}\) (time slot~1) and \(\{f_1,f_3,f_4\}\) (time slot~2).
The four coverage constraints, expressed in binary-assignment form, are
\begin{align}
%\mathcal{Y}_1&=\{x_{1,A},0,0,0,0,0,0,0,0,0,p\}, \notag \\
%\mathcal{Y}_2&=\{0,x_{2,A}, x_{2,B}, x_{2,C},0,0,0,0,0,0,p\}, \notag  \\
%\mathcal{Y}_3&=\{0,0,0,0,x_{3,A}, x_{3,B}, x_{3,C},0,0,0,p\}, \notag  \\
%\mathcal{Y}_4&=\{0,0,0,0,0,0,0,x_{4,A}, x_{4,B}, x_{4,C},p\}, \notag 
\mathcal{Y}_1&=\{x_{1,A}\}, \notag \\
\mathcal{Y}_2&=\{x_{2,A}, x_{2,B}, x_{2,C}\}, \notag  \\
\mathcal{Y}_3&=\{x_{3,A}, x_{3,B}, x_{3,C}\}, \notag  \\
\mathcal{Y}_4&=\{x_{4,A}, x_{4,B}, x_{4,C}\}, \notag 
\end{align}
%in which element 0 is retained in the above sets, as it will subsequently be mapped to the corresponding gadgets defined earlier. 
and the six slot-conflict constraints, expressed in binary-assignment form, are
\begin{align}
%\mathcal{Y}_5&=\{x_{1,A}, x_{2,A},0,0,x_{3,A},0,0,0,0,0,P\}, \notag  \\
%\mathcal{Y}_6&=\{0,0,x_{2,B},0,0,x_{3,A},0,0,0,0,P\}, \notag  \\
%\mathcal{Y}_7&=\{0,0,0,x_{2,C},0,0,x_{3,C},0,0,0,P\}, \notag  \\
%\mathcal{Y}_8&=\{x_{1,A},0,0,0,x_{3,A},0,0,x_{4,A},0,0,P\}, \notag  \\
%\mathcal{Y}_9&=\{0,0,0,0,0,x_{3,B},0,0,x_{4,B},0,P\}, \notag  \\
%\mathcal{Y}_{10}&=\{0,0,0,0,0,0,x_{3,C},0,0,x_{4,C},P\}. \notag 
\mathcal{Y}_5&=\{x_{1,A}, x_{2,A},x_{3,A},P\}, \notag  \\
\mathcal{Y}_6&=\{x_{2,B},x_{3,A},P\}, \notag  \\
\mathcal{Y}_7&=\{x_{2,C},x_{3,C},P\}, \notag  \\
\mathcal{Y}_8&=\{x_{1,A},x_{3,A},x_{4,A},P\}, \notag  \\
\mathcal{Y}_9&=\{x_{3,B},x_{4,B},P\}, \notag  \\
\mathcal{Y}_{10}&=\{x_{3,C},x_{4,C},P\}. \notag 
\end{align}
%Combining these two sets of constraints gives us the constraint satisfaction problem for the case under consideration, depicted in the left panel of Fig. \ref{fig:binxor1_minimal_0} in which non-participating elements are kept (denoted by $0$) as we need them to build gadgets depicted in Fig. \ref{fig:additional} to realize correct synchorinization between the $xor_1$ gadgets.
Combining these two sets of constraints yields the constraint satisfaction problem shown in the left panel of Fig.~\ref{fig:binxor1_minimal_0}. Non-participating elements in each constraint (denoted by $0$) are retained, as they are required to construct the gadgets in Fig.~\ref{fig:additional} that ensure proper synchronization between the $xor_1$ gadgets.

\begin{figure*}
\centering
\begin{minipage}{0.36\linewidth}
\begingroup
\fontsize{7pt}{7pt}\selectfont
\begin{lstlisting}
flight: 1     2       3       4    
gate:   A   A B C   A B C   A B C
                                   dummy
$\mathcal{Y}_1:$        1   0 0 0   0 0 0   0 0 0    p
$\mathcal{Y}_2:$        0   1 1 1   0 0 0   0 0 0    p
$\mathcal{Y}_3:$        0   0 0 0   1 1 1   0 0 0    p
$\mathcal{Y}_4:$        0   0 0 0   0 0 0   1 1 1    p
$\mathcal{Y}_5:$        1   1 0 0   1 0 0   0 0 0    P
$\mathcal{Y}_6:$        0   0 1 0   0 1 0   0 0 0    P
$\mathcal{Y}_7:$        0   0 0 1   0 0 1   0 0 0    P
$\mathcal{Y}_8:$        1   0 0 0   1 0 0   1 0 0    P
$\mathcal{Y}_9:$        0   0 0 0   0 1 0   0 1 0    P
$\mathcal{Y}_{10}:$        0   0 0 0   0 0 1   0 0 1    P
\end{lstlisting}
\endgroup 
\end{minipage}
\hfill
\begin{minipage}{0.3\linewidth}
\begingroup
\fontsize{7pt}{7pt}\selectfont
\begin{lstlisting}
flight:  2     3     4   
gate:   B C   B C   B C
                        dummy
$\mathcal{Y}_2:$        1 1   0 0   0 0   p
$\mathcal{Y}_3:$        0 0   1 1   0 0   P
$\mathcal{Y}_4:$        0 0   0 0   1 1   p
$\mathcal{Y}_6:$        1 0   1 0   0 0   P
$\mathcal{Y}_7:$        0 1   0 1   0 0   P
$\mathcal{Y}_9:$        0 0   1 0   1 0   P
$\mathcal{Y}_{10}:$        0 0   0 1   0 1   P
\end{lstlisting}
\endgroup 
\end{minipage}
\hfill
\begin{minipage}{0.3\linewidth}
\begingroup
\fontsize{7pt}{7pt}\selectfont
\begin{lstlisting} 
flight:  2     3     4   
gate:   B C   B C   B C
                        dummy
$\mathcal{Y}_{10}:$        0 0   0 1   0 1   P
$\mathcal{Y}_4:$        0 0   0 0   1 1   p
$\mathcal{Y}_9:$        0 0   1 0   1     P
$\mathcal{Y}_7:$        0 1   0 1         P
$\mathcal{Y}_3:$        0 0   1 1         p
$\mathcal{Y}_6:$        1 0   1           P
$\mathcal{Y}_2:$        1 1               p
\end{lstlisting}
\endgroup
\end{minipage}
\caption{Constraint formulation in binary assignment form for gate assignment problem with three gates and four flights: variables involved in a constraint are denoted by 1 (otherwise 0). Presence of dummy variables is indicated by a trailing uppercase `P` (otherwise `p`). The left panel shows the initial CSP to be solved. In the middle panel, the algorithm described in Fig.~\ref{fig:multi_xor1_proc} is applied to eliminate singles and reduce the complexity of the optimization problem, and subsequently, reduce atom counts later used for RAA arrangement. The right panel presents a rearranged constraint set, minimizing the number of atoms required for implementation on RAA platforms. This final arrangement in the right panel serves as the basis for constructing the RAA configuration shown in Fig.~\ref{fig:RAAA_minimal_4}.}
\label{fig:binxor1_minimal_0}
\end{figure*}

\begin{figure*}
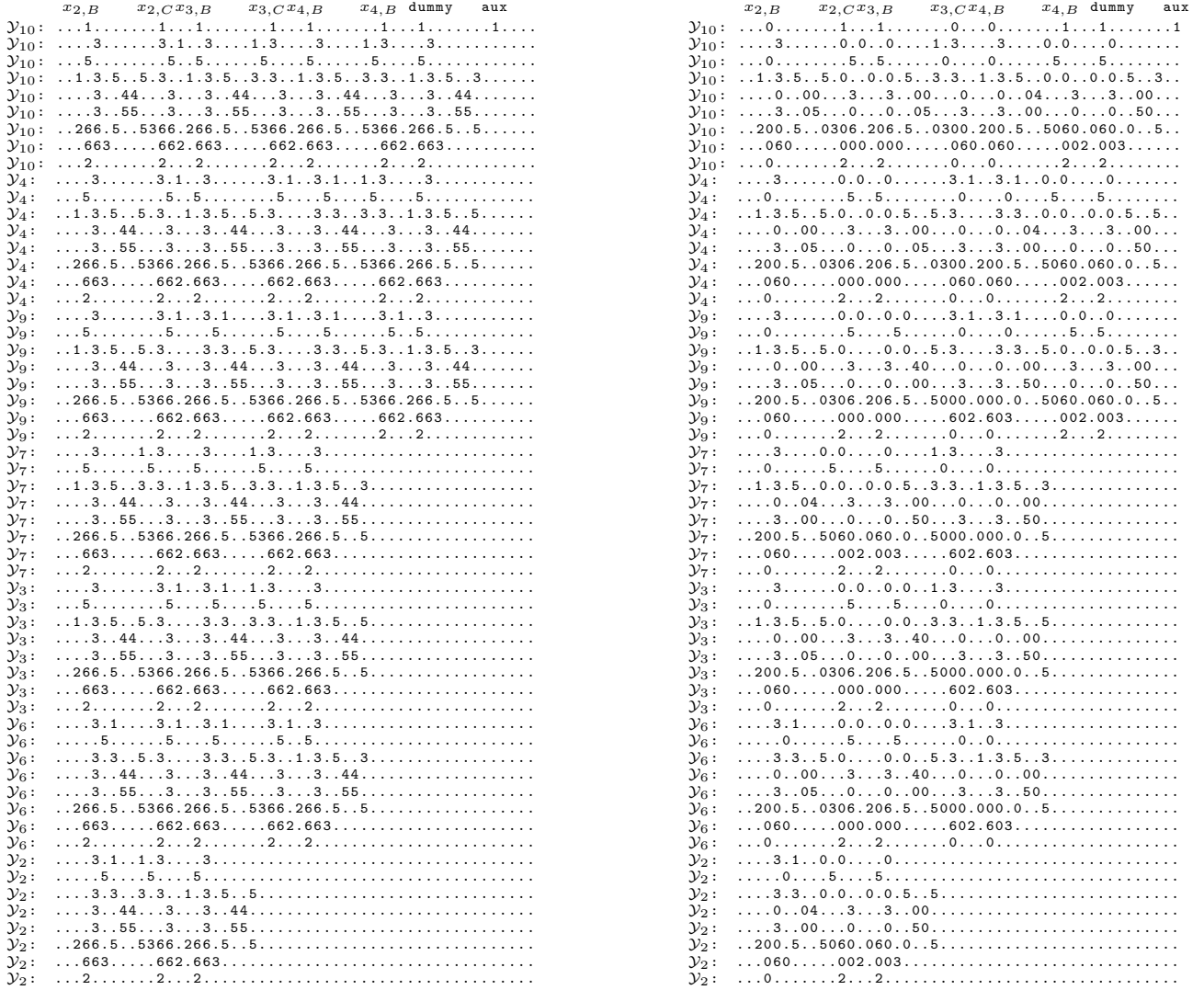

\centering
    
\begin{minipage}{0.45\textwidth}
\begingroup
\fontsize{6pt}{7pt}\selectfont
\begin{lstlisting}
      $x_{2,B}$        $x_{2,C}$$x_{3,B}$            $x_{3,C}$$x_{4,B}$            $x_{4,B}$ dummy       aux 
$\mathcal{Y}_{10}$: ...1.......1...1.......1...1.......1...1.......1....
$\mathcal{Y}_{10}$: ....3......3.1..3....1.3....3....1.3....3...........
$\mathcal{Y}_{10}$: ...5........5..5......5....5......5....5............
$\mathcal{Y}_{10}$: ..1.3.5..5.3..1.3.5..3.3..1.3.5..3.3..1.3.5..3......
$\mathcal{Y}_{10}$: ....3..44...3...3..44...3...3..44...3...3..44.......
$\mathcal{Y}_{10}$: ....3..55...3...3..55...3...3..55...3...3..55.......
$\mathcal{Y}_{10}$: ..266.5..5366.266.5..5366.266.5..5366.266.5..5......
$\mathcal{Y}_{10}$: ...663.....662.663.....662.663.....662.663..........
$\mathcal{Y}_{10}$: ...2.......2...2.......2...2.......2...2............
$\mathcal{Y}_4$:    ....3......3.1..3......3.1..3.1..1.3....3...........
$\mathcal{Y}_4$:    ...5........5..5........5....5....5....5............
$\mathcal{Y}_4$:    ..1.3.5..5.3..1.3.5..5.3....3.3..3.3..1.3.5..5......
$\mathcal{Y}_4$:    ....3..44...3...3..44...3...3..44...3...3..44.......
$\mathcal{Y}_4$:    ....3..55...3...3..55...3...3..55...3...3..55.......
$\mathcal{Y}_4$:    ..266.5..5366.266.5..5366.266.5..5366.266.5..5......
$\mathcal{Y}_4$:    ...663.....662.663.....662.663.....662.663..........
$\mathcal{Y}_4$:    ...2.......2...2.......2...2.......2...2............
$\mathcal{Y}_9$:    ....3......3.1..3.1....3.1..3.1....3.1..3...........
$\mathcal{Y}_9$:    ...5........5....5......5....5......5..5............
$\mathcal{Y}_9$:    ..1.3.5..5.3....3.3..5.3....3.3..5.3..1.3.5..3......
$\mathcal{Y}_9$:    ....3..44...3...3..44...3...3..44...3...3..44.......
$\mathcal{Y}_9$:    ....3..55...3...3..55...3...3..55...3...3..55.......
$\mathcal{Y}_9$:    ..266.5..5366.266.5..5366.266.5..5366.266.5..5......
$\mathcal{Y}_9$:    ...663.....662.663.....662.663.....662.663..........
$\mathcal{Y}_9$:    ...2.......2...2.......2...2.......2...2............
$\mathcal{Y}_7$:    ....3....1.3....3....1.3....3.......................
$\mathcal{Y}_7$:    ...5......5....5......5....5........................
$\mathcal{Y}_7$:    ..1.3.5..3.3..1.3.5..3.3..1.3.5..3..................
$\mathcal{Y}_7$:    ....3..44...3...3..44...3...3..44...................
$\mathcal{Y}_7$:    ....3..55...3...3..55...3...3..55...................
$\mathcal{Y}_7$:    ..266.5..5366.266.5..5366.266.5..5..................
$\mathcal{Y}_7$:    ...663.....662.663.....662.663......................
$\mathcal{Y}_7$:    ...2.......2...2.......2...2........................
$\mathcal{Y}_3$:    ....3......3.1..3.1..1.3....3.......................
$\mathcal{Y}_3$:    ...5........5....5....5....5........................
$\mathcal{Y}_3$:    ..1.3.5..5.3....3.3..3.3..1.3.5..5..................
$\mathcal{Y}_3$:    ....3..44...3...3..44...3...3..44...................
$\mathcal{Y}_3$:    ....3..55...3...3..55...3...3..55...................
$\mathcal{Y}_3$:    ..266.5..5366.266.5..5366.266.5..5..................
$\mathcal{Y}_3$:    ...663.....662.663.....662.663......................
$\mathcal{Y}_3$:    ...2.......2...2.......2...2........................
$\mathcal{Y}_6$:    ....3.1....3.1..3.1....3.1..3.......................
$\mathcal{Y}_6$:    .....5......5....5......5..5........................
$\mathcal{Y}_6$:    ....3.3..5.3....3.3..5.3..1.3.5..3..................
$\mathcal{Y}_6$:    ....3..44...3...3..44...3...3..44...................
$\mathcal{Y}_6$:    ....3..55...3...3..55...3...3..55...................
$\mathcal{Y}_6$:    ..266.5..5366.266.5..5366.266.5..5..................
$\mathcal{Y}_6$:    ...663.....662.663.....662.663......................
$\mathcal{Y}_6$:    ...2.......2...2.......2...2........................
$\mathcal{Y}_2$:    ....3.1..1.3....3...................................
$\mathcal{Y}_2$:    .....5....5....5....................................
$\mathcal{Y}_2$:    ....3.3..3.3..1.3.5..5..............................
$\mathcal{Y}_2$:    ....3..44...3...3..44...............................
$\mathcal{Y}_2$:    ....3..55...3...3..55...............................
$\mathcal{Y}_2$:    ..266.5..5366.266.5..5..............................
$\mathcal{Y}_2$:    ...663.....662.663..................................
$\mathcal{Y}_2$:    ...2.......2...2....................................
\end{lstlisting}
\endgroup
\end{minipage}
\hfill
\begin{minipage}{0.45\textwidth}
\begingroup
\fontsize{6pt}{7pt}\selectfont
\begin{lstlisting}
      $x_{2,B}$        $x_{2,C}$$x_{3,B}$            $x_{3,C}$$x_{4,B}$            $x_{4,B}$ dummy       aux 
$\mathcal{Y}_{10}$: ...0.......1...1.......0...0.......1...1.......1
$\mathcal{Y}_{10}$: ....3......0.0..0....1.3....3....0.0....0.......
$\mathcal{Y}_{10}$: ...0........5..5......0....0......5....5........
$\mathcal{Y}_{10}$: ..1.3.5..5.0..0.0.5..3.3..1.3.5..0.0..0.0.5..3..
$\mathcal{Y}_{10}$: ....0..00...3...3..00...0...0..04...3...3..00...
$\mathcal{Y}_{10}$: ....3..05...0...0..05...3...3..00...0...0..50...
$\mathcal{Y}_{10}$: ..200.5..0306.206.5..0300.200.5..5060.060.0..5..
$\mathcal{Y}_{10}$: ...060.....000.000.....060.060.....002.003......
$\mathcal{Y}_{10}$: ...0.......2...2.......0...0.......2...2........
$\mathcal{Y}_4$:    ....3......0.0..0......3.1..3.1..0.0....0.......
$\mathcal{Y}_4$:    ...0........5..5........0....0....5....5........
$\mathcal{Y}_4$:    ..1.3.5..5.0..0.0.5..5.3....3.3..0.0..0.0.5..5..
$\mathcal{Y}_4$:    ....0..00...3...3..00...0...0..04...3...3..00...
$\mathcal{Y}_4$:    ....3..05...0...0..05...3...3..00...0...0..50...
$\mathcal{Y}_4$:    ..200.5..0306.206.5..0300.200.5..5060.060.0..5..
$\mathcal{Y}_4$:    ...060.....000.000.....060.060.....002.003......
$\mathcal{Y}_4$:    ...0.......2...2.......0...0.......2...2........
$\mathcal{Y}_9$:    ....3......0.0..0.0....3.1..3.1....0.0..0.......
$\mathcal{Y}_9$:    ...0........5....5......0....0......5..5........
$\mathcal{Y}_9$:    ..1.3.5..5.0....0.0..5.3....3.3..5.0..0.0.5..3..
$\mathcal{Y}_9$:    ....0..00...3...3..40...0...0..00...3...3..00...
$\mathcal{Y}_9$:    ....3..05...0...0..00...3...3..50...0...0..50...
$\mathcal{Y}_9$:    ..200.5..0306.206.5..5000.000.0..5060.060.0..5..
$\mathcal{Y}_9$:    ...060.....000.000.....602.603.....002.003......
$\mathcal{Y}_9$:    ...0.......2...2.......0...0.......2...2........
$\mathcal{Y}_7$:    ....3....0.0....0....1.3....3...................
$\mathcal{Y}_7$:    ...0......5....5......0....0....................
$\mathcal{Y}_7$:    ..1.3.5..0.0..0.0.5..3.3..1.3.5..3..............
$\mathcal{Y}_7$:    ....0..04...3...3..00...0...0..00...............
$\mathcal{Y}_7$:    ....3..00...0...0..50...3...3..50...............
$\mathcal{Y}_7$:    ..200.5..5060.060.0..5000.000.0..5..............
$\mathcal{Y}_7$:    ...060.....002.003.....602.603..................
$\mathcal{Y}_7$:    ...0.......2...2.......0...0....................
$\mathcal{Y}_3$:    ....3......0.0..0.0..1.3....3...................
$\mathcal{Y}_3$:    ...0........5....5....0....0....................
$\mathcal{Y}_3$:    ..1.3.5..5.0....0.0..3.3..1.3.5..5..............
$\mathcal{Y}_3$:    ....0..00...3...3..40...0...0..00...............
$\mathcal{Y}_3$:    ....3..05...0...0..00...3...3..50...............
$\mathcal{Y}_3$:    ..200.5..0306.206.5..5000.000.0..5..............
$\mathcal{Y}_3$:    ...060.....000.000.....602.603..................
$\mathcal{Y}_3$:    ...0.......2...2.......0...0....................
$\mathcal{Y}_6$:    ....3.1....0.0..0.0....3.1..3...................
$\mathcal{Y}_6$:    .....0......5....5......0..0....................
$\mathcal{Y}_6$:    ....3.3..5.0....0.0..5.3..1.3.5..3..............
$\mathcal{Y}_6$:    ....0..00...3...3..40...0...0..00...............
$\mathcal{Y}_6$:    ....3..05...0...0..00...3...3..50...............
$\mathcal{Y}_6$:    ..200.5..0306.206.5..5000.000.0..5..............
$\mathcal{Y}_6$:    ...060.....000.000.....602.603..................
$\mathcal{Y}_6$:    ...0.......2...2.......0...0....................
$\mathcal{Y}_2$:    ....3.1..0.0....0...............................
$\mathcal{Y}_2$:    .....0....5....5................................
$\mathcal{Y}_2$:    ....3.3..0.0..0.0.5..5..........................
$\mathcal{Y}_2$:    ....0..04...3...3..00...........................
$\mathcal{Y}_2$:    ....3..00...0...0..50...........................
$\mathcal{Y}_2$:    ..200.5..5060.060.0..5..........................
$\mathcal{Y}_2$:    ...060.....002.003..............................
$\mathcal{Y}_2$:    ...0.......2...2................................
\end{lstlisting}
\endgroup
\end{minipage}
\caption{The ASCII representation of the RAA arrangement for the minimal example, with ``.'' indicating unoccupied sites and numbers corresponding to vertices. The left panel shows the initial $xor_1$ atom array configuration, obtained from the right panel of Fig.~\ref{fig:binxor1_minimal_0}, while the right panel presents the activation pattern with active vertices marked by nonzero numbers and inactive ones with zeros. This pattern corresponds to the ground state of the RAA arrangement, which represents the solution to the CSP. In an experimental setup, this solution can be determined by checking only the activity of the vertices within the $xor_1$ units, without the need to inspect all vertices (and hence all atoms).}
\label{fig:RAAA_minimal_4}
\end{figure*}

It is straightforward to pair the variables in each constraint and build up their $xor_1$ gadgets according to the composite structures given in Figs. \ref{fig:Cross-Copy} and \ref{fig:additional}. The ground state of the resulting structure would be the solution to our gate assignment problem. In practice, though, it is possible to reduce the number of required vertices even further. This can be done via two methods: the algorithm given in Fig. \ref{fig:multi_xor1_proc} and reordering the vertices and constraints. We use these two methods chronologically.

We first need to find the singles, which in our case is the first constraint in the left panel of Fig. \ref{fig:binxor1_minimal_0}, i.e. $\mathcal{Y}_1=\{x_{1,A}\}$. This tells us that using $xor_1$, we find $x_{1A}=1$ and subsequently, we have $x_{2,A}=x_{3,A}=x_{4,A}=0$ in $\mathcal{Y}_5$ and $\mathcal{Y}_8$, which enables us to remove these optimization variables and their corresponding constraints. This cleaning procedure reduces our problem to one given in the middle panel of Fig. \ref{fig:binxor1_minimal_0}. Next, we reorder the optimization variables by their occurrence. After that, the sequence of constraints is re-ordered so that some decision variables are only used at the beginning. This allows us to cut off all synchronization connections afterwards, resulting in a configuration depicted in the right panel of Fig. \ref{fig:binxor1_minimal_0} which has considerably lower atom counts compared to the initial case of the RAA configuration. Although this step may be classified as preprocessing, it incurs negligible computational overhead and thus has no practical impact on the overall preprocessing cost.

The final binary assignment (right panel of Fig. \ref{fig:binxor1_minimal_0}) is translated into an RAA configuration using composite structures, with the resulting arrangement presented in Fig. \ref{fig:RAAA_minimal_4} in which we employ an ASCII representation rather than a graphical depiction for better readability. This RAA configuration is then optimized to obtain MWIS using our developed MIP solver (see Appendix Sec. \ref{sec:gateadd} for more details), yielding the arrangement shown in Fig. \ref{fig:RAAA_minimal_4}. The corresponding activation pattern, specifically the optimization vertices in each $xor_1$ units, provides the final solution to our optimization problem, summarized in Table \ref{tab:minimal-example-gate-allocations}. In Appendix Sec. \ref{sec:gateadd}, we present additional examples and their corresponding RAA configurations, further illustrating the generality of the approach. Appendix Sec. \ref{sec:nqueens} also considers the $N$-queens problem, an $NP$-hard optimization problem that can be naturally implemented on Rydberg platforms using our proposed 
$xor_1$ gadget.

\begin{table}
\centering
\caption{Flight-Gate-Allocations from RAA activations. Only gate A can serve wide-body planes. }
\label{tab:minimal-example-gate-allocations}
% First subtable: Slots 1-8
\begin{tabular}{c|cccc}
\toprule
\textbf{Gate} & \multicolumn{4}{c}{\textbf{Time Slots}} \\
& \textbf{0} & \textbf{1} & \textbf{2} & \textbf{3} \\
\midrule

A     & $f_1$  & $f_1$  & $f_1$  &     \\
B     &     & $f_3$  & $f_3$  &     \\
C     & $f_2$  & $f_2$  & $f_4$  & $f_4$  \\
\bottomrule
\end{tabular}
\end{table}

\section{Conclusion} \label{sec:conclusion}
In this work, we demonstrated that constraint satisfaction problems can be encoded on Rydberg platforms using our $xor_1$ gadgets that are both physically native and experimentally scalable. The clique-based perspective adopted here clarifies how combinatorial constraints can be enforced directly through geometric layout and Rydberg  blockade interactions, rather than through large energy penalties or extensive classical preprocessing.

The proposed $xor_1$ gadget provides a compact and flexible mechanism for implementing exactly-one constraints within this setting, enabling problem encodings that respect the limited connectivity and control resources of current RAAs. Beyond the specific examples considered, the framework suggests a general strategy for translating constrained satisfaction problems into analog quantum dynamics using minimal overhead.

More broadly, our results highlight the potential of RAAs as programmable analog quantum optimizers in which problem structure is embedded directly into many-body interactions. Beyond traditional scheduling and allocation tasks, closely related constraint-driven optimization problems arise in molecular science and drug discovery, including molecular docking and related combinatorial design problems that involve steric exclusion and pattern compatibility constraints \cite{garrigues2025scalableheuristicmoleculardocking, PhysRevResearch.6.043020}. Similar combinatorial challenges also appear in materials science, such as designing metamaterials, where discrete structural choices must satisfy geometric and physical constraints \cite{PhysRevResearch.2.013319}. As neutral-atom hardware continues to scale in system size and control fidelity, such physically motivated encodings may provide a viable route toward exploring optimization regimes that remain intractable for classical approaches.

\section{Acknowledgments}
The authors gratefully acknowledge the funding provided by the Hamburgische Investitions- und Förderbank for the project "Efficient Quantum Algorithms for Aviation".
SP and DJ acknowledge support from the Hamburg Quantum Computing Initiative (HQIC) project EFRE. 
The EFRE project is co-financed by ERDF of the European Union and by “Fonds of the Hamburg Ministry of Science, Research, Equalities and Districts (BWFGB)”.
DJ acknowledges support from the Cluster of Excellence `Advanced Imaging of Matter' of the Deutsche Forschungsgemeinschaft (DFG) - EXC 2056 - project ID 390715994, the European Union’s Horizon Programme (HORIZON-CL42021-
DIGITALEMERGING-02-10) Grant Agreement 101080085 QCFD, DFG project ‘Quantencomputing mit neutralen Atomen’ 
(JA 1793/1-1, Japan-JST-DFG-ASPIRE 2024), and funding from the Federal Ministry of Research, Technology and Space (BMFTR) under the grant BeRyQC.

\begin{widetext}
\renewcommand{\theequation}{A.\arabic{equation}}
\setcounter{equation}{0}

\section{Appendix}

\renewcommand{\thefigure}{A\arabic{figure}}
\setcounter{figure}{0}

%\section{A. Blockade mechanism and limitations in Rydberg platforms}

\subsection{ Gate assignment problem: Larger illustrations} \label{sec:gateadd}

In this section, we study larger problem instances as case studies. Specifically, we consider three representative problem sizes, classified as small, medium, and large (depicted by examples I, II, and III, respectively) according to the number of flights, gates, and time slots. To keep solver runtimes manageable, we use three data sets with gradually increasing complexity. We use our developed MIP solver to show that the ground state of our RAA arrangement is the solution to our optimization problem. The solver translates the constraints to sets, which are then used to build RAA arrangements. The MWIS (the resulting ILP) is solved using PuLP library and HiGHS solver.  

The small instance comprises three gates, seven time slots, and five arriving flights. One of the gates is restricted to narrow-body aircraft, and two of the five flights are operated by wide-body aircraft.
The medium instance comprises three gates, fourteen time slots, and ten arriving flights. Again, one gate is restricted to narrow-body aircraft, while five of the ten flights are operated by wide-body aircraft.
The large instance comprises five gates, fifteen time slots, and nineteen arriving flights. Two gates are restricted to narrow-body aircraft, and six of the nineteen flights are operated by wide-body aircraft. Table \ref{tab:example-sizes} summarizes details about these three cases.  Following the same procedure as described in the previous section, we find the sets of constraints as depicted in Figs. \ref{fig:binxor1_example_2} and \ref{fig:binxor1_example_3}.

\begin{table}[htbp]
\centering
\caption{The three examples of the gate assignment problem and their details.}
\label{tab:example-sizes}
\begin{tabular}{cccccc}
\toprule
\textbf{gates} & \textbf{slots} & \textbf{flights} & \textbf{constraints}  & 
\makecell{\textbf{narrow} \\ \textbf{gates}} & 
\makecell{\textbf{wbody} \\ \textbf{planes}} \\
\midrule
3 & 7 & 5 & 9 & 1 & 2 \\
3 & 10 & 10 & 18 & 1 & 4 \\
5 & 15 & 19 & 27 & 2 & 6 \\
\bottomrule
\end{tabular}
\end{table}

\begin{figure}[htbp]
\begin{minipage}{0.45\textwidth}
\centering
\begin{lstlisting}
$\mathcal{Y}_{01}$: 00 00 00 11 00 00 10 0p
$\mathcal{Y}_{02}$: 00 00 11 00 00 01 0p
$\mathcal{Y}_{03}$: 00 00 00 00 00 11 0P
$\mathcal{Y}_{04}$: 00 11 00 00 00 10 0p
$\mathcal{Y}_{05}$: 00 00 00 01 01 0P
$\mathcal{Y}_{06}$: 00 00 00 00 11 0p
$\mathcal{Y}_{07}$: 01 00 00 01 01 0P
$\mathcal{Y}_{08}$: 00 00 00 11 0P
$\mathcal{Y}_{09}$: 10 00 00 11 0P
$\mathcal{Y}_{10}$: 01 00 01 0P
$\mathcal{Y}_{11}$: 01 01 01 0P
$\mathcal{Y}_{12}$: 10 00 10 0P
$\mathcal{Y}_{13}$: 10 10 10 0P
$\mathcal{Y}_{14}$: 01 01 0P
$\mathcal{Y}_{15}$: 10 10 0P
$\mathcal{Y}_{16}$: 11 0p
\end{lstlisting}
\end{minipage}
\hfill
\begin{minipage}{0.45\textwidth}
\begingroup
\fontsize{5pt}{7pt}\selectfont
\begin{lstlisting}
$\mathcal{Y}_{01}$: 00 01 00 00 00 00 00 01 00 00 00 00 01 0P
$\mathcal{Y}_{02}$: 00 00 00 00 00 00 00 00 00 00 00 00 11 0p
$\mathcal{Y}_{03}$: 00 10 00 00 00 00 00 10 00 00 00 00 10 0P
$\mathcal{Y}_{04}$: 00 00 00 00 00 00 00 00 00 00 01 11 0p
$\mathcal{Y}_{05}$: 00 01 00 00 00 00 00 01 00 00 00 01 0P
$\mathcal{Y}_{06}$: 00 10 00 00 00 00 00 10 00 00 00 10 0P
$\mathcal{Y}_{07}$: 00 00 00 00 00 00 00 00 00 00 11 0P
$\mathcal{Y}_{08}$: 00 00 00 00 00 00 00 11 00 00 10 0p
$\mathcal{Y}_{09}$: 00 00 00 00 00 00 00 00 11 0P
$\mathcal{Y}_{10}$: 00 00 00 00 00 00 11 00 01 0p
$\mathcal{Y}_{11}$: 00 00 00 00 11 00 00 00 10 0p
$\mathcal{Y}_{12}$: 00 00 00 11 00 00 00 01 0p
$\mathcal{Y}_{13}$: 00 00 00 00 00 00 00 11 0P
$\mathcal{Y}_{14}$: 00 00 11 00 00 00 00 10 0p
$\mathcal{Y}_{15}$: 00 00 00 00 01 01 01 0P
$\mathcal{Y}_{16}$: 00 01 00 00 00 00 01 0P
$\mathcal{Y}_{17}$: 00 00 00 00 10 10 10 0P
$\mathcal{Y}_{18}$: 00 10 00 00 00 00 10 0P
$\mathcal{Y}_{19}$: 00 00 00 00 00 11 0p
$\mathcal{Y}_{20}$: 01 00 00 00 01 01 0P
$\mathcal{Y}_{21}$: 10 00 00 00 11 0P
$\mathcal{Y}_{22}$: 01 00 01 01 0P
$\mathcal{Y}_{23}$: 01 00 00 01 0P
$\mathcal{Y}_{24}$: 10 00 00 10 0P
$\mathcal{Y}_{25}$: 01 00 01 0P
$\mathcal{Y}_{26}$: 10 00 10 0P
$\mathcal{Y}_{27}$: 00 11 0p
$\mathcal{Y}_{28}$: 11 0p
\end{lstlisting}
\endgroup
\end{minipage}
\caption{Pairwise mapping of decision variables obtained from Table \ref{tab:example-sizes} for constraints of example I and II for the gate assignment problem.}
\label{fig:binxor1_example_2}
\end{figure}

\begin{figure*}[htbp]
\centering
\begingroup
\fontsize{4pt}{5pt}\selectfont
\begin{lstlisting}
$\mathcal{Y}_{1}$:  00 00 00 00 00 00 00 00 00 00 00 00 00 00 00 00 00 00 00 00 00 00 00 00 00 00 00 00 00 00 11 10 00 00 00 00 00 00 00 01 10 0p
$\mathcal{Y}_{2}$:  00 00 00 00 00 00 00 00 00 00 00 00 00 00 00 00 00 00 00 00 00 00 00 00 00 00 00 00 00 00 00 00 00 00 00 00 00 00 00 10 10 0P
$\mathcal{Y}_{3}$:  00 00 00 00 00 00 00 00 00 00 00 00 00 00 00 00 00 00 00 00 00 00 00 00 00 00 00 00 00 00 00 00 00 00 00 00 00 00 01 01 0P
$\mathcal{Y}_{4}$:  00 00 00 00 00 00 00 00 00 00 00 00 00 00 00 00 00 00 00 00 00 00 00 00 00 00 00 00 01 11 00 00 00 00 00 00 00 00 01 10 0p
$\mathcal{Y}_{5}$:  00 00 00 00 00 00 00 00 00 00 00 00 00 00 00 00 00 00 00 00 00 00 00 00 11 10 00 00 00 00 00 00 00 00 00 00 00 01 10 0p
$\mathcal{Y}_{6}$:  00 00 00 00 00 00 00 00 00 00 00 00 00 00 00 00 00 00 00 01 00 00 00 00 00 00 00 00 00 00 00 00 00 00 00 00 00 00 10 0P
$\mathcal{Y}_{7}$:  00 00 00 00 00 00 00 00 00 00 00 00 00 00 00 00 00 00 00 10 00 00 00 00 00 00 00 00 00 00 00 00 00 00 00 00 00 01 0P
$\mathcal{Y}_{8}$:  00 00 00 00 00 01 00 00 00 00 00 00 00 00 00 00 00 00 00 00 00 10 00 00 00 00 00 00 10 00 00 00 00 00 01 00 00 10 0P
$\mathcal{Y}_{9}$:  00 00 00 00 00 00 00 00 00 00 00 00 00 00 00 00 00 00 00 00 00 00 00 00 00 00 00 00 00 00 00 00 00 00 00 11 11 10 0p
$\mathcal{Y}_{10}$: 00 00 00 00 00 10 00 00 00 00 00 00 00 00 00 00 00 00 00 00 01 00 00 00 00 00 00 01 00 00 00 00 00 00 10 00 01 0P
$\mathcal{Y}_{11}$: 00 00 00 00 01 00 00 00 00 00 00 00 00 00 00 00 00 00 00 00 10 00 00 00 00 00 00 10 00 00 00 00 00 01 00 00 10 0P
$\mathcal{Y}_{12}$: 00 00 00 00 00 00 00 00 00 00 00 00 00 00 00 00 00 00 00 01 00 00 00 00 00 00 00 00 00 00 00 00 00 10 00 01 0P
$\mathcal{Y}_{13}$: 00 00 00 00 00 00 00 00 00 00 00 00 00 00 00 00 00 00 00 10 00 00 00 00 00 00 00 00 00 00 00 00 01 00 00 10 0P
$\mathcal{Y}_{14}$: 00 00 00 00 00 00 00 00 00 00 00 00 00 00 00 00 00 00 00 00 00 00 00 00 00 00 00 00 00 00 00 00 01 11 11 0p
$\mathcal{Y}_{15}$: 00 00 00 00 00 00 00 00 00 00 00 00 00 00 10 00 00 00 00 00 00 00 10 00 00 00 00 00 00 00 00 00 10 0P
$\mathcal{Y}_{16}$: 00 00 00 00 00 00 00 00 00 00 00 00 00 00 00 00 00 01 11 00 00 00 00 00 00 00 00 00 00 00 00 01 10 0p
$\mathcal{Y}_{17}$: 00 00 00 00 00 00 00 00 00 00 00 00 00 01 00 00 00 00 00 00 00 01 00 00 00 00 00 00 00 00 00 01 0P
$\mathcal{Y}_{18}$: 00 00 00 00 00 00 00 00 00 00 00 00 00 00 00 00 00 00 00 00 00 00 00 00 00 00 00 00 00 01 00 10 0P
$\mathcal{Y}_{19}$: 00 00 00 00 00 00 00 10 00 00 00 00 00 00 00 00 00 00 00 00 00 00 00 00 00 00 00 00 00 01 00 10 0P
$\mathcal{Y}_{20}$: 00 00 00 00 00 00 00 00 00 00 00 00 00 00 00 00 00 00 00 00 00 00 00 00 00 00 00 00 00 10 10 0P
$\mathcal{Y}_{21}$: 00 00 00 00 00 00 01 00 00 00 00 00 00 00 00 00 00 00 00 00 00 00 00 00 00 00 00 00 00 10 10 0P
$\mathcal{Y}_{22}$: 00 00 00 00 00 00 10 00 00 00 00 00 00 00 00 00 00 00 00 00 00 00 00 00 00 00 00 00 01 10 0P
$\mathcal{Y}_{23}$: 00 00 00 00 00 00 00 00 00 00 00 00 00 00 00 00 00 00 00 00 00 00 00 00 00 00 00 00 01 10 0P
$\mathcal{Y}_{24}$: 00 00 00 00 00 00 00 00 00 00 00 00 00 00 00 00 00 00 00 00 00 00 00 00 00 00 00 11 10 0p
$\mathcal{Y}_{25}$: 00 00 00 00 00 01 00 10 00 00 00 00 00 00 00 00 00 00 00 00 00 10 00 00 00 10 00 00 10 0P
$\mathcal{Y}_{26}$: 00 00 00 00 00 10 01 00 00 00 00 00 00 00 00 00 00 00 00 00 01 00 00 00 01 00 00 01 0P
$\mathcal{Y}_{27}$: 00 00 00 00 01 00 10 00 00 00 00 00 00 00 00 00 00 00 00 00 10 00 00 00 10 00 00 10 0P
$\mathcal{Y}_{28}$: 00 00 00 00 00 00 00 00 00 00 00 00 00 00 00 00 00 00 00 00 00 00 00 00 00 01 11 0p
$\mathcal{Y}_{29}$: 00 00 00 00 00 00 00 00 00 00 00 00 00 00 00 01 00 10 01 00 00 00 00 01 00 00 01 0P
$\mathcal{Y}_{30}$: 00 00 00 00 00 01 00 00 00 00 00 00 00 00 00 00 00 00 01 00 00 00 00 00 00 00 01 0P
$\mathcal{Y}_{31}$: 00 00 00 00 00 00 00 00 00 00 00 00 00 00 00 10 01 00 10 00 00 00 01 00 00 01 0P
$\mathcal{Y}_{32}$: 00 00 00 00 00 10 00 00 00 00 00 00 00 00 00 00 00 00 10 00 00 00 00 00 00 01 0P
$\mathcal{Y}_{33}$: 00 00 00 00 01 00 00 00 00 00 00 00 00 00 00 00 00 01 00 00 00 00 00 00 01 0P
$\mathcal{Y}_{34}$: 00 00 00 00 00 00 00 00 00 00 00 00 00 00 01 00 10 01 00 00 00 01 00 00 01 0P
$\mathcal{Y}_{35}$: 00 00 00 00 00 00 00 10 00 00 00 00 00 00 00 00 00 00 00 00 00 00 00 01 0P
$\mathcal{Y}_{36}$: 00 00 00 00 00 00 01 00 00 00 00 00 00 00 00 00 00 00 00 00 00 00 00 10 0P
$\mathcal{Y}_{37}$: 00 00 00 00 00 00 10 00 00 00 00 00 00 00 00 00 00 00 00 00 00 00 01 0P
$\mathcal{Y}_{38}$: 00 10 00 00 00 00 00 00 00 00 00 00 00 00 00 01 00 10 00 00 00 00 10 0P
$\mathcal{Y}_{39}$: 00 00 00 00 00 00 00 00 00 00 00 00 00 00 00 00 00 00 00 00 11 11 10 0p
$\mathcal{Y}_{40}$: 01 00 00 00 00 00 00 00 00 00 00 00 00 00 00 10 01 00 00 00 00 01 0P
$\mathcal{Y}_{41}$: 10 00 00 00 00 00 00 00 00 00 00 00 00 00 01 00 10 00 00 00 00 10 0P
$\mathcal{Y}_{42}$: 00 00 00 00 00 00 00 00 00 00 00 00 00 00 10 00 00 00 00 00 01 0P
$\mathcal{Y}_{43}$: 00 00 00 00 00 00 00 00 00 00 00 00 00 01 00 00 00 00 00 00 10 0P
$\mathcal{Y}_{44}$: 00 00 00 00 00 00 00 00 00 00 00 00 00 00 00 00 00 01 11 11 0p
$\mathcal{Y}_{45}$: 00 00 00 00 00 00 00 00 00 00 00 00 00 00 00 00 11 10 0p
$\mathcal{Y}_{46}$: 00 00 00 00 00 00 00 00 00 00 00 00 00 01 11 11 0p
$\mathcal{Y}_{47}$: 00 10 01 00 00 00 00 00 00 01 00 00 00 10 0P
$\mathcal{Y}_{48}$: 00 00 00 00 00 00 00 00 00 00 00 11 11 10 0p
$\mathcal{Y}_{49}$: 00 10 01 00 10 00 00 00 00 01 00 00 00 10 0P
$\mathcal{Y}_{50}$: 01 00 10 01 00 00 00 00 00 10 00 00 01 0P
$\mathcal{Y}_{51}$: 01 00 10 00 00 00 00 00 00 10 00 00 01 0P
$\mathcal{Y}_{52}$: 10 01 00 10 00 00 00 00 01 00 00 00 10 0P
$\mathcal{Y}_{53}$: 10 01 00 00 00 00 00 00 01 00 00 00 10 0P
$\mathcal{Y}_{54}$: 00 00 00 00 00 00 00 00 10 00 00 01 0P
$\mathcal{Y}_{55}$: 00 00 00 00 00 00 00 00 10 00 01 01 0P
$\mathcal{Y}_{56}$: 00 00 00 00 00 00 00 01 00 00 10 10 0P
$\mathcal{Y}_{57}$: 00 00 00 00 00 00 00 01 00 00 00 10 0P
$\mathcal{Y}_{58}$: 00 00 00 00 00 00 00 00 10 00 01 0P
$\mathcal{Y}_{59}$: 00 00 00 11 10 00 00 00 00 00 11 0p
$\mathcal{Y}_{60}$: 00 00 00 00 00 00 00 01 00 00 10 0P
$\mathcal{Y}_{61}$: 00 00 00 00 00 00 00 01 11 11 0p
$\mathcal{Y}_{62}$: 00 10 01 00 10 00 00 00 01 0P
$\mathcal{Y}_{63}$: 01 00 10 01 00 00 00 00 10 0P
$\mathcal{Y}_{64}$: 10 01 00 10 00 00 00 01 0P
$\mathcal{Y}_{65}$: 00 00 00 00 00 00 11 10 0p
$\mathcal{Y}_{66}$: 00 00 00 00 01 11 0p
$\mathcal{Y}_{67}$: 00 00 01 00 10 0P
$\mathcal{Y}_{68}$: 00 00 10 01 0P
$\mathcal{Y}_{69}$: 00 01 00 10 0P
$\mathcal{Y}_{70}$: 00 01 11 0p
$\mathcal{Y}_{71}$: 11 10 0p
\end{lstlisting}
\endgroup
\caption{Pairwise mapping of decision variables obtained from Table \ref{tab:example-sizes} for constraints of example III.}
\label{fig:binxor1_example_3}
\end{figure*}

Due to space limitations, we do not present the detailed RAA layouts for these three examples, nor the corresponding activation patterns that encode the solutions to the associated optimization problems. 
%Instead, these layouts and activation patterns are provided as companion files (see \ref{??}). 
The resulting activation patterns on the RAAs yield solutions summarized in Tables \ref{tab:example-gate-allocations-1}, \ref{tab:example-gate-allocations-2}, and \ref{tab:benchmark3-gate-allocations}, confirming the applicability of our proposed framework and $xor_1$ gadget across different problem sizes. A detailed comparison of resource requirements for each example is presented in Table \ref{tab:raaa_benchmark} and contrasted with the corresponding QUBO-based implementations.

We find that the optimized $xor_1$ gadget requires up to 37\% fewer atoms than the QUBO formulation. Moreover, the 
$xor_1$ gadget enforces constraints through geometric blockade interactions rather than energetic penalty terms, implying that the required Rydberg detuning remains independent of problem size. In contrast, QUBO-based encodings rely on penalty terms to impose constraints, leading to detuning requirements that scale with the magnitude of the QUBO coefficients and, consequently, with problem complexity. Evidently, the QUBO requires additional preprocessing for weight assignment for each atom in the RAA arrangement, which is not needed for our $xor_1$ gadget.

\begin{table}[htbp]
\centering
\caption{Results for flight allocations of example I: 3 gates, 5 flights, and 7 time-slots. Gate A can only serve narrow-body machines.}
\label{tab:example-gate-allocations-1}
\begin{tabular}{@{}c|ccccccc@{}}
\toprule
\textbf{Gate} & \multicolumn{7}{c}{\textbf{Time Slots}} \\
& \textbf{1} & \textbf{2} & \textbf{3} & \textbf{4} & \textbf{5} & \textbf{6} & \textbf{7} \\
\midrule
A &  $f_1$ & $f_1$ & $f_1$ & -- & $f_3$ & $f_3$ & $f_3$ \\
B   &  -- & $f_4$ & $f_4$ & $f_4$ & $f_4$ & -- & -- \\
C   &  -- & -- & $f_2$ & $f_2$ & $f_5$ & $f_5$ & $f_5$ \\
\bottomrule
\end{tabular}
\end{table}

\begin{table}[htbp]
\centering
\caption{Results for flight allocations of example II: 3 gates, 10 flights, and 14 time-slots. Gate A can only serve narrow-body machines.}
\label{tab:example-gate-allocations-2}
\begin{tabular}{@{}c|ccccccccccccccccc@{}}
\toprule
\textbf{Gate} & \multicolumn{14}{c}{\textbf{Time Slots}} \\
& \textbf{1} & \textbf{2} & \textbf{3} & \textbf{4} & \textbf{5} & \textbf{6} & \textbf{7} & \textbf{8} & \textbf{9} & \textbf{10} & \textbf{11} & \textbf{12} & \textbf{13} & \textbf{14} \\
\midrule
A &  $f_1$ & $f_1$ & $f_1$ & -- & $f_3$ & $f_3$ & $f_3$ & -- & $f_7$ & $f_7$ & $f_7$  & --  & --  & -- \\
B   &  -- & $f_4$ & $f_4$ & $f_4$ & $f_4$ & $f_6$ & $f_6$ & $f_6$ & $f_9$ & $f_9$ & $f_{10}$ & $f_{10}$ & $f_{10}$ & $f_{10}$ \\
C   &  -- & -- & $f_2$ & $f_2$ & $f_5$ & $f_5$ & $f_5$ & $f_8$ & $f_8$ & $f_8$ & $f_8$  & --  & -- & -- \\
\bottomrule
\end{tabular}
\end{table}

 \begin{table*}[htbp]
\centering
\caption{Results for flight allocations of example III: 19 flights, 5 gates, 16 slots. Gates in parentheses are the only narrow-body ones.}
\label{tab:benchmark3-gate-allocations}

\begin{tabular}{c|cccccccccccccccc}
\toprule
\textbf{Gate} & \multicolumn{16}{c}{\textbf{Time Slots}} \\
& \textbf{1} & \textbf{2} & \textbf{3} & \textbf{4} & \textbf{5} & \textbf{6} & \textbf{7} & \textbf{8}
& \textbf{9} & \textbf{10} & \textbf{11} & \textbf{12} & \textbf{13} & \textbf{14} & \textbf{15} & \textbf{16} \\
\midrule
(A) & $f_1$  & $f_1$  & $f_1$  &  --   & $f_3$  & $f_3$  & $f_3$  &  --   & $f_7$  & $f_7$  & $f_7$  &  --   &   --  &   --  &  --   &   --  \\
B   &   --  & $f_4$  & $f_4$  & $f_4$  & $f_4$  & $f_6$  & $f_6$  & $f_6$  & $f_9$  & $f_9$  & $f_{10}$ & $f_{10}$ & $f_{10}$ & $f_{10}$ &   --  &   -- \\
C   &  --   &   --  & $f_2$  & $f_2$  & $f_5$  & $f_5$  & $f_5$  & $f_8$  & $f_8$  & $f_8$  & $f_8$  & $f_8$  &  --   &   --  &   --  &   --  \\
D   & $f_{11}$ & $f_{11}$ & $f_{11}$ & $f_{11}$ &   --  & $f_{16}$ & $f_{16}$ & $f_{16}$ & $f_{17}$ & $f_{17}$ & $f_{17}$ &   --  & $f_{18}$ & $f_{18}$ & $f_{18}$ &  --   \\
(E) &   --  & $f_{12}$ & $f_{12}$ & $f_{12}$ & $f_{13}$ & $f_{13}$ & $f_{13}$ &   --  & $f_{14}$ & $f_{14}$ & $f_{15}$ & $f_{15}$ & $f_{19}$ & $f_{19}$ & $f_{19}$ &  --   \\
\bottomrule
\end{tabular}
\end{table*}

\begin{table}[h!]
\centering
\begin{tabular}{l|c c c}
\toprule
\textbf{} & \textbf{example I} & \textbf{example II} & \textbf{example III} \\
\midrule
\textbf{gates} & 3 &  3 & 5 \\
\textbf{slots} & 7 & 10 & 15 \\
\textbf{flights} & 5 & 10 & 20 \\
\textbf{opt. vars.} & 13 & 26 & 81 \\
\textbf{\(xor_1\)-atoms} & 2693 & 8119  & 57124 \\
\textbf{\(xor_1\)-atoms*} & 3781 & 13219 & 98203 \\
\textbf{$\Delta$\(xor_1\)} & 0.29 & 0.39 & 0.42 \\
\textbf{QUBO-atoms} & 3 076 & 10 298 & 89 696 \\
\textbf{\(xor_1\)-RAAA} & $86 \times 131$ & $170 \times 235$ & $494 \times 571$ \\
\textbf{QUBO-RAAA} & $108 \times 108$ & $200 \times 200$ & $596 \times 596$ \\
\textbf{\(xor_1\)-detuning} & 6 & 6 & 6 \\
\textbf{QUBO-detuning} & 416 & 1320 & 9156 \\
\textbf{objective} & 4238 & 13126 & 94341 \\
\bottomrule
\end{tabular}
\caption{Results of the $xor_1$-based Rydberg atom array arrangements (RAAAs) for the three gate-assignment example scenarios introduced in Table~\ref{tab:example-sizes}. The increasing number of optimization variables (opt. vars.) reflects the growing number of flights to be assigned to gates under time-slot and geometric constraints. The quantities $xor_1$-atoms$^*$ and $xor_1$-atoms denote the number of Rydberg atoms required to implement the array before and after the optimization procedure, respectively, while $\Delta xor_1$ gives the fractional reduction in atom count
achieved by this optimization. For comparison, QUBO-atoms indicates the number of atoms required to implement the corresponding RAA using the QUBO-based UnitDiskMapping.jl approach from Ref. \cite{nguyen2023quantum}. $xor_1$- and QUBO-detuning are the maximum range of Rydberg atom detuning needed for the RAAA to enforce feasible solutions/satisfy all constraints. Detuning values are reported in dimensionless units normalized by the characteristic Rabi frequency $\Omega_0$ \cite{nguyen2023quantum}, which we set to $\Omega_0 = 1$ for convenience. The entries labeled $xor_1$- and QUBO-RAAA specify the spatial dimensions of the resulting RAAs. The objective value corresponds to the weighted sum obtained from the ground-state solution of each RAA.
}
\label{tab:raaa_benchmark}
\end{table}

%\begin{table}[h!]
%\centering
%\begin{tabular}{c|rrrr}
%\hline
%\textbf{$\lambda$} & \multicolumn{4}{c}{\textbf{Rydberg detuning}} \\
%\cline{2-5}
%& \textbf{min} & \textbf{max} & \textbf{mean} & \textbf{var} \\
%\hline
%10 & -26 & 34 & 2.9 & 5.8 \\
%100 & -296 & 304 & 2.9 & 481 \\
%1 000 & -2 996 & 3 004 & 2.9 & 48 075 \\
%\hline
%\end{tabular}
%\caption{Rydberg-detuning due to different choices of penalty factor $(\lambda)$ for standard QUBO approach.}
%\label{tab:rydberg_detuning}
%\end{table}

\
subsection{ Realization of the $N$-queens problem via $xor_1$ gadget
} \label{sec:nqueens}

The \(N\)-queens problem is a canonical constraint-satisfaction problem in combinatorial optimization and computer science. It asks whether \(N\) queens can be placed on an \(N \times N\) chessboard such that no two queens attack each other, \emph{i.e.}, no two queens share the same row, column, or diagonal. While the decision version of the problem admits polynomial-time solutions for fixed \(N\), its formulation as a constraint satisfaction problem with local exclusion rules provides a natural benchmark for hardware-efficient quantum encodings. Generalized variants of the problem, including additional constraints or colored queens, are known to be $NP$-hard \cite{GentEtAl2017,Torggler2019quantumnqueens}.

In this work, we focus on the feasibility (constraint-satisfaction) version of the problem rather than the optimization variant that maximizes the number of non-attacking queens. Each possible board position is associated with a binary decision variable \(q_{x,y} \in \{0,1\}\), indicating whether a queen occupies position \((x,y)\), with \(0 \le x < N\) and \(0 \le y < N\).
The constraints enforce that at most one queen may occupy any row, column, or diagonal. The mathematical formulations of these constraints are given as follows
\begin{equation}
    \sum_{i=0}^{x+i \leq N} q_{x+i, y} \; \leq 1, \;\;\; \forall \; (0 \leq y < N),
    \label{queensrow}
\end{equation}

\begin{equation}
    \sum_{i=0}^{y+i \leq N} q_{x, y+i} \; \leq 1, \;\;\; \forall \; (0 \leq x < N),
    \label{queenscolumn}
\end{equation}

\begin{equation}
    \sum_{i=1}^{\substack{ x+i \leq N \\y+i \leq N}} q_{x+i, y+i} \; \leq 1, \;\;\; 
    \forall x,y \in 
    \left\{ \substack{ x=0,y=0 \\ y=0, 0 < x < N-1 \\ x=0, 0 < y < N-1} \right\},
    \label{queensdiagup}
\end{equation}

\begin{equation}
    \sum_{i=1}^{\substack{ x+i \leq N \\y-i > 0}} q_{x+i, y-i} \; \leq 1, \;\;\; 
    \forall x,y \in 
    \left\{ \substack{ x=0, y=N-1 \\ x=0, 0 < y < N-1 \\ y=N-1, 0 < x < N-1} \right\},
    \label{queensdiagdn}
\end{equation}
which results in the total number of constraints that scales linearly with system size, and is given by
\begin{equation}
    N_{cstr.} = 2 N + 2 \left( 2 (N - 2) + 1 \right)  = 6 (N - 1)
    \label{queens_num_constr}
\end{equation}
corresponding to row constraints (Eq.~\eqref{queensrow} with Fig.~\ref{fig:queens_constr}a), column constraints (Eq.~\eqref{queenscolumn} with Fig.~\ref{fig:queens_constr}b), upward diagonals (Eq.~\eqref{queensdiagup} with Fig.~\ref{fig:queens_constr}c), and downward diagonals (Eq.~\eqref{queensdiagdn} with Fig.~\ref{fig:queens_constr}d

\begin{figure*}
\centering
\includegraphics[width=0.95\linewidth]{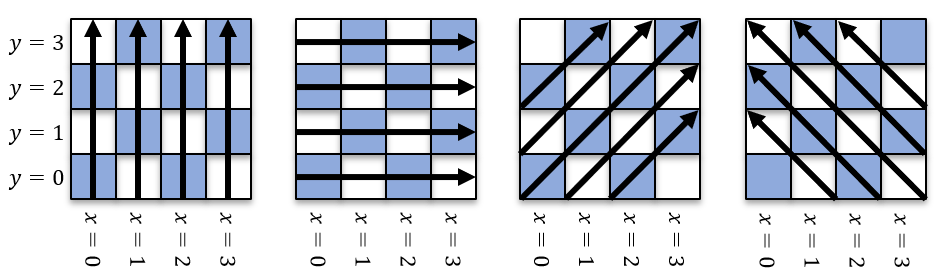}
\llap{\parbox[b]{12.6in}{(a)\\\rule{0ex}{1.88in}}}
\llap{\parbox[b]{9.46in}{(b)\\\rule{0ex}{1.88in}}}
\llap{\parbox[b]{6.32in}{(c)\\\rule{0ex}{1.88in}}}
\llap{\parbox[b]{3.14in}{(d)\\\rule{0ex}{1.88in}}}
\caption{Example of queens problem on a chessboard: sketch of constraint-pattern for four cases of at-most-one queen on a row (a), on a column (b), on diagonals upwards (c), and on diagonals downwards (d). The graphical structures are then translated to the subsets given in Fig. \ref{queens_RAAA_bin_assign_00}. }
\label{fig:queens_constr}
\end{figure*}

\begin{figure*}
\centering
\begin{minipage}{0.5\linewidth}
\begingroup
\fontsize{6pt}{6pt}\selectfont
\begin{lstlisting}
Column:   0         1         2         3
Row:   0 1 2 3   0 1 2 3   0 1 2 3   0 1 2 3
                                             dummy  Constraint
$\mathcal{Y}_1:$       1 1 1 1   0 0 0 0   0 0 0 0   0 0 0 0   P    row 0
$\mathcal{Y}_2:$       0 0 0 0   1 1 1 1   0 0 0 0   0 0 0 0   P    row 1
$\mathcal{Y}_3:$       0 0 0 0   0 0 0 0   1 1 1 1   0 0 0 0   P    row 2
$\mathcal{Y}_4:$       0 0 0 0   0 0 0 0   0 0 0 0   1 1 1 1   P    row 3
$\mathcal{Y}_5:$       1 0 0 0   1 0 0 0   1 0 0 0   1 0 0 0   P    col. 0
$\mathcal{Y}_6:$       0 1 0 0   0 1 0 0   0 1 0 0   0 1 0 0   P    col. 1
$\mathcal{Y}_7:$       0 0 1 0   0 0 1 0   0 0 1 0   0 0 1 0   P    col. 2
$\mathcal{Y}_8:$       0 0 0 1   0 0 0 1   0 0 0 1   0 0 0 1   P    col. 3
$\mathcal{Y}_9:$       1 0 0 0   0 1 0 0   0 0 1 0   0 0 0 1   P    diag. up. 0
$\mathcal{Y}_{10}:$       0 1 0 0   0 0 1 0   0 0 0 1   0 0 0 0   P    diag. up. 1
$\mathcal{Y}_{11}:$       0 0 1 0   0 0 0 1   0 0 0 0   1 0 0 0   P    diag. up. 2
$\mathcal{Y}_{12}:$       0 0 0 0   1 0 0 0   0 1 0 0   0 0 1 0   P    diag. up. 3
$\mathcal{Y}_{13}:$       0 0 0 0   0 0 0 0   1 0 0 0   0 1 0 0   P    diag. up. 4
$\mathcal{Y}_{14}:$       0 0 0 1   0 0 1 0   0 1 0 0   1 0 0 0   P    diag. up. 0
$\mathcal{Y}_{15}:$       0 0 0 0   0 0 0 1   0 0 1 0   0 1 0 0   P    diag. up. 1
$\mathcal{Y}_{16}:$       0 0 0 0   0 0 0 0   0 0 0 1   0 0 1 0   P    diag. up. 2
$\mathcal{Y}_{17}:$       0 0 1 0   0 1 0 0   1 0 0 0   0 0 0 0   P    diag. up. 3
$\mathcal{Y}_{18}:$       0 1 0 0   1 0 0 0   0 0 0 0   0 0 0 0   P    diag. up. 4
\end{lstlisting}
\endgroup
\end{minipage}
\hfill
\begin{minipage}{0.45\linewidth}
\begingroup
\fontsize{6pt}{6pt}\selectfont
\begin{lstlisting} 
Column:   0         1         2         3
Row:   0 1 2 3   0 1 2 3   0 1 2 3   0 1 2 3
                                             dummy
$\mathcal{Y}_4:$       0 0 0 0   0 0 0 0   0 0 0 0   1 1 1 1   P                                            
$\mathcal{Y}_8:$       0 0 0 1   0 0 0 1   0 0 0 1   0 0 0 1   P
$\mathcal{Y}_9:$       1 0 0 0   0 1 0 0   0 0 1 0   0 0 0 1   P
$\mathcal{Y}_7:$       0 0 1 0   0 0 1 0   0 0 1 0   0 0 1 0   P
$\mathcal{Y}_{12}:$       0 0 0 0   1 0 0 0   0 1 0 0   0 0 1 0   P
$\mathcal{Y}_{16}:$       0 0 0 0   0 0 0 0   0 0 0 1   0 0 1 0   P
$\mathcal{Y}_6:$       0 1 0 0   0 1 0 0   0 1 0 0   0 1       P
$\mathcal{Y}_{13}:$       0 0 0 0   0 0 0 0   1 0 0 0   0 1       P
$\mathcal{Y}_{15}:$       0 0 0 0   0 0 0 1   0 0 1 0   0 1       P
$\mathcal{Y}_5:$       1 0 0 0   1 0 0 0   1 0 0 0   1 0       P
$\mathcal{Y}_{11}:$       0 0 1 0   0 0 0 1   0 0 0 0   1 0       P
$\mathcal{Y}_{14}:$       0 0 0 1   0 0 1 0   0 1 0 0   1 0       P
$\mathcal{Y}_3:$       0 0 0 0   0 0 0 0   1 1 1 1             P
$\mathcal{Y}_{10}:$       0 1 0 0   0 0 1 0   0 0 0 1             P
$\mathcal{Y}_{17}:$       0 0 1 0   0 1 0 0   1 0                 P
$\mathcal{Y}_2:$       0 0 0 0   1 1 1 1                       P
$\mathcal{Y}_{18}:$       0 1 0 0   1 0                           P
$\mathcal{Y}_1:$       1 1 1 1                                 P
\end{lstlisting}
\endgroup
\end{minipage}
\caption{Example of queens problem on a $4 \times\ 4$ chessboard: Binary mapping of decision variables to $xor_1$-constraints analogues to the procedure described earlier and shown in Fig. \ref{fig:binxor1_minimal_0}. Number of constraints: left panel, the initial state of the constraints, and right panel, the re-ordered analogues to the  procedure described earlier. The re-ordered version is used for creating the RAAA, which is shown in Fig. \ref{fig:queens_RAAA} and its activation pattern is given in Fig. \ref{fig:queens_RAAA_activation}. }
\label{queens_RAAA_bin_assign_00}
\end{figure*}

Each of these constraints can be naturally implemented using the proposed \(xor_1\) gadget, which enforces exactly-one or at-most-one selection rule through geometric blockade interactions. 
%This makes the \(N\)-queens problem particularly well suited for realization on Rydberg atom arrays, where row, column, and diagonal conflicts correspond directly to local blockade constraints in the interaction graph.
Combining all of the constraints provides us with the optimization problem that we need to solve which, for a $4 \times 4$ example, it is given in Fig. \ref{queens_RAAA_bin_assign_00}. Using the $xor_1$ gadgets introduced in Fig. \ref{fig:Cross-Copy} and \ref{fig:additional}, we can obtain the RAAA given in Fig. \ref{fig:queens_RAAA}. The activation pattern of this RAAA is given in Fig. \ref{fig:queens_RAAA_activation}, which corresponds to one of the solutions of the $4 \times 4$ queens problem, confirming the validity of our gadget for this class of optimization problems. 

Furthermore, we investigate $N$-queens instances of varying system sizes and provide a detailed comparison of the corresponding resource requirements. The results for each instance are summarized in Table~\ref{tab:queens_benchmark} and contrasted with the associated QUBO-based implementations from Ref. \cite{nguyen2023quantum}.

We find that for sufficiently large problem instances (see Table~\ref{tab:queens_benchmark} for details), both the optimized and non-optimized $xor_1$ gadget require substantially fewer atoms (up to 54\% fewer) than the corresponding QUBO formulation. Moreover, the detuning required by the $xor_1$ gadget remains constant and is approximately 99\% smaller than the detuning required in the QUBO-based encoding, which increases with problem complexity. As in the gate-assignment problem discussed in the previous section, this demonstrates that the proposed $xor_1$ gadget does not require problem-dependent preprocessing to determine individual atom detunings, in contrast to QUBO-based methods. This is evident in Table \ref{tab:preptimes}.

\begin{table}[h!]
  \centering
  \begin{tabular}{@{}l|c c c c c c}
    \toprule
    \textbf{complexity} & \multicolumn{6}{c}{\textbf{board-sizes}} \\
    \textbf{\& resources} & $ 4 \times 4 $ & $ 8 \times 8 $ & $ 12 \times 12 $
    & $ 16 \times 16$  & $ 20 \times 20 $ & $ 40 \times 40 $ \\
    \midrule
    \textbf{constraints} & 18 & 42 & 66 & 90 & 114 & 234\\
    \textbf{opt. vars.}    & 16 & 64 & 144 & 256 & 400 & 1600 \\
    \textbf{\(xor_1\)-atoms}          & 4238 & 35390 & 122606 & 295262 & 582734 & 4770494\\
    \textbf{\(xor_1\)-atoms*}         & 5292 & 46644 & 163100 & 393828 & 777996 & 6371316\\
    \textbf{$\Delta$\(xor_1\)-atoms} & 0.20  & 0.24   & 0.25 & 0.25 & 0.25 & 0.25\\
    \textbf{QUBO-atoms}          & 1118 & 16766 & 83806 & 216144 & 642398 & 10265616\\
    \textbf{\(xor_1\)-RAAA}           & $ 112 \times 145 $ 
    & $ 400 \times 337 $ 
    & $ 800 \times 529 $ 
    & $ 1 552 \times  724$
    & $ 2 416 \times 916 $ 
    & $ 9 616 \times 1 876 $\\
    \textbf{QUBO-RAAA} & $ 64 \times 64 $ 
    & $256 \times 256 $ 
    & $576 \times 576 $ 
    & $ 1 024 \times 1 024 $
    & $ 1 600 \times 1 600 $
    & $ 6 400 \times 6 400 $\\
    \textbf{\(xor_1\)-detuning} & 6 & 6 & 6 & 6 & 6 & 6\\        
    \textbf{QBUBO-detuning} & 324 & 2772 & 9636 & 23220 & 45828 & 374868 \\
    \bottomrule
  \end{tabular}
  \caption{Results of the $xor_1$-based Rydberg atom array arrangements (RAAAs) for the queens-placement-problem for six board sizes. 
    The quantities $xor_1$-atoms$^*$ and $xor_1$-atoms denote the number of Rydberg atoms required to implement the array before and after the optimization procedure, respectively, while $\Delta xor_1$ gives the fractional reduction in atom count achieved by this optimization. 
    The problems consist solely of constraints of form $ \sum_{x_i \in \mathcal{Y}_k} x_i \leq 1 $ and therefore due to no singles, only re-orderings, re-arrangements, and tail-cuts are possible.
    For comparison, QUBO-atoms indicates the number of atoms required to implement the corresponding RAA using the QUBO-based UnitDiskMapping.jl approach. The entries labeled $xor_1$- and QUBO-RAAA specify the spatial dimensions of the resulting RAAs. 
    $xor_1$- and QUBO-detuning are the maximum range of Rydberg atom detuning needed for the RAAA to enforce feasible solutions/satisfy all constraints. Detuning values are reported in dimensionless units normalized by the characteristic Rabi frequency $\Omega_0$ \cite{nguyen2023quantum}, which we set to $\Omega_0 = 1$ for convenience.
  }
  \label{tab:queens_benchmark}
\end{table}

\begin{table}[h!]
  \centering
  \begin{tabular}{@{}l|c c c c c c}
    \toprule
    \textbf{MWIS \(\rightarrow\) RAAA} & \multicolumn{6}{c}{\textbf{board-sizes}} \\  
    \textbf{pre-processing times} & $ 4 \times 4 $ & $ 8 \times 8 $ & $ 12 \times 12 $
                         & $ 16 \times 16$  & $ 20 \times 20 $ & $ 40 \times 40 $ \\
    \midrule
    \textbf{opt. vars.}           & 16     & 64     & 144    & 256    & 400    & 1600 \\
    \textbf{UnitDiskMapping.jl abs. [s]}   & 3.9E1  & 1.0E1  & 3.2E1  & 9.2E1  & 2.0E2  & 3.1E3 \\
    \textbf{UnitDiskMapping.jl rel. [s]}   & 2.4E-1 & 1.6E-1 & 2.2E-1 & 3.5E-1 & 5.0E-1 & 1.9E0 \\
    \textbf{\(xor_1\) binary mapping abs. [s]} & 4.0E-4 & 3.0E-3 & 1.4E-2 & 3.6E-2 & 7.5E-2 & 9.7E-1 \\
    \textbf{\(xor_1\) binary mapping rel. [s]} & 2.0E-5 & 5.0E-5 & 1.0E-4 & 1.4E-4 & 1.9E-4 & 6.0E-4 \\
    \midrule
    \textbf{ratio rel. times} (UDM / \(xor_1\)) & 1.2E4 & 3.2E3 & 2.2E3 & 2.5E3 & 3.5E3 & 3.2E3 \\
    \bottomrule
  \end{tabular}
  \caption{Comparison of preprocessing times for UnitDiskMapping.jl \cite{nguyen2023quantum} and the proposed \(xor_1\) binary-mapping method when constructing RAAs from MWIS-encoded \(N\)-queens instances. 
Reported times are given both in absolute terms (total preprocessing time per instance) and in relative terms (time per optimization variable). 
For UnitDiskMapping.jl, the greedy (non-optimal) preprocessing variant was employed, as the optimal preprocessing routine exceeded the imposed time limit (\(10^5\)~s) for board sizes larger than \(4\times4\). 
The ratio of preprocessing time per decision variable remains approximately constant across problem sizes, indicating similar scaling behavior; however, the \(xor_1\) method is more than three orders of magnitude (\(>10^3\times\)) faster.
}
\label{tab:preptimes}
\end{table}

\begin{figure}[htbp]
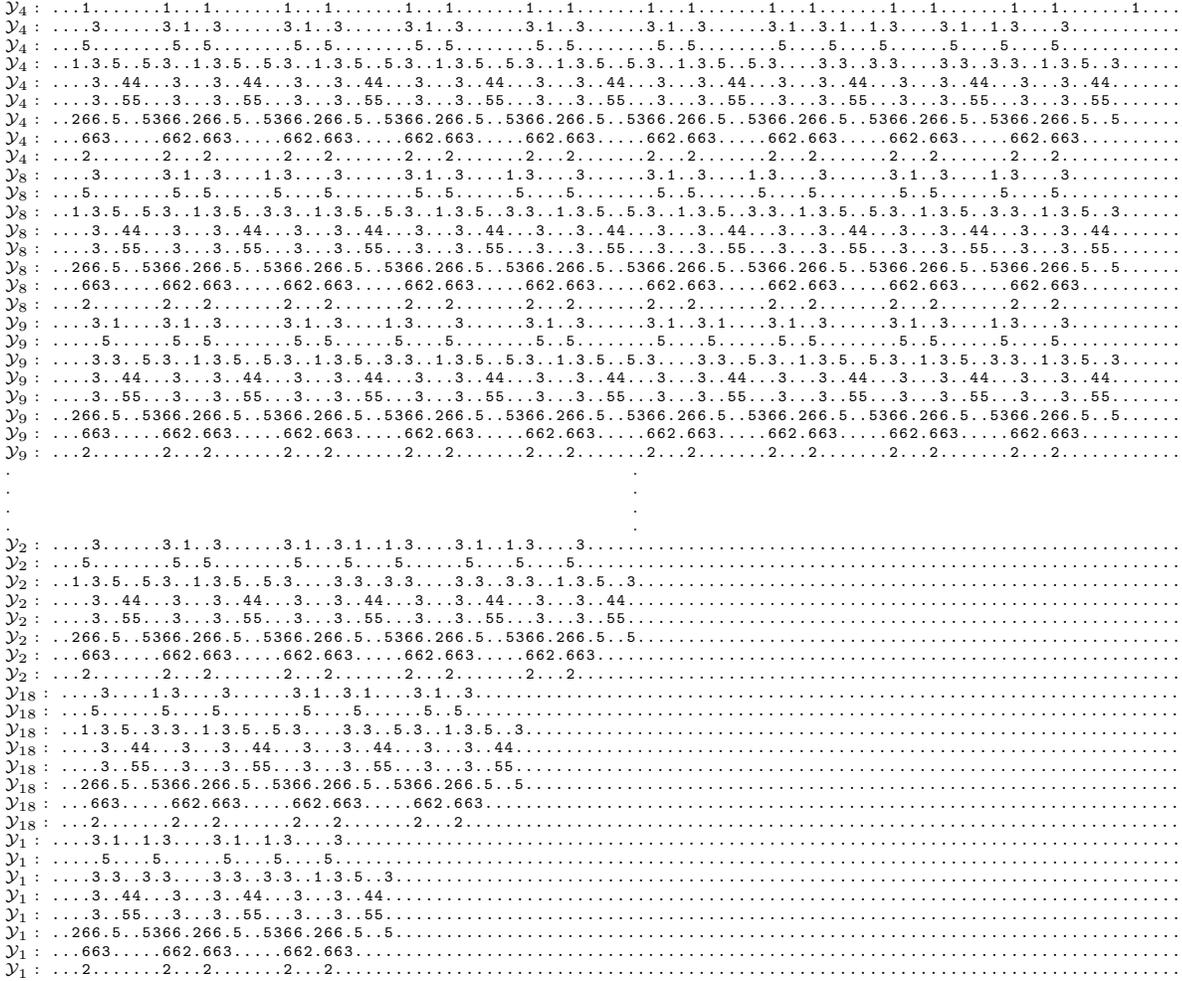

    \centering
    \begingroup
    \fontsize{6pt}{7pt}\selectfont
    \begin{lstlisting}
$\mathcal{Y}_{4}:$ ...1.......1...1.......1...1.......1...1.......1...1.......1...1.......1...1.......1...1.......1...1.......1....
$\mathcal{Y}_{4}:$ ....3......3.1..3......3.1..3......3.1..3......3.1..3......3.1..3......3.1..3.1..1.3....3.1..1.3....3...........
$\mathcal{Y}_{4}:$ ...5........5..5........5..5........5..5........5..5........5..5........5....5....5......5....5....5............
$\mathcal{Y}_{4}:$ ..1.3.5..5.3..1.3.5..5.3..1.3.5..5.3..1.3.5..5.3..1.3.5..5.3..1.3.5..5.3....3.3..3.3....3.3..3.3..1.3.5..3......
$\mathcal{Y}_{4}:$ ....3..44...3...3..44...3...3..44...3...3..44...3...3..44...3...3..44...3...3..44...3...3..44...3...3..44.......
$\mathcal{Y}_{4}:$ ....3..55...3...3..55...3...3..55...3...3..55...3...3..55...3...3..55...3...3..55...3...3..55...3...3..55.......
$\mathcal{Y}_{4}:$ ..266.5..5366.266.5..5366.266.5..5366.266.5..5366.266.5..5366.266.5..5366.266.5..5366.266.5..5366.266.5..5......
$\mathcal{Y}_{4}:$ ...663.....662.663.....662.663.....662.663.....662.663.....662.663.....662.663.....662.663.....662.663..........
$\mathcal{Y}_{4}:$ ...2.......2...2.......2...2.......2...2.......2...2.......2...2.......2...2.......2...2.......2...2............
$\mathcal{Y}_{8}:$ ....3......3.1..3....1.3....3......3.1..3....1.3....3......3.1..3....1.3....3......3.1..3....1.3....3...........
$\mathcal{Y}_{8}:$  ...5........5..5......5....5........5..5......5....5........5..5......5....5........5..5......5....5............
$\mathcal{Y}_{8}:$  ..1.3.5..5.3..1.3.5..3.3..1.3.5..5.3..1.3.5..3.3..1.3.5..5.3..1.3.5..3.3..1.3.5..5.3..1.3.5..3.3..1.3.5..3......
$\mathcal{Y}_{8}:$  ....3..44...3...3..44...3...3..44...3...3..44...3...3..44...3...3..44...3...3..44...3...3..44...3...3..44.......
$\mathcal{Y}_{8}:$  ....3..55...3...3..55...3...3..55...3...3..55...3...3..55...3...3..55...3...3..55...3...3..55...3...3..55.......
$\mathcal{Y}_{8}:$  ..266.5..5366.266.5..5366.266.5..5366.266.5..5366.266.5..5366.266.5..5366.266.5..5366.266.5..5366.266.5..5......
$\mathcal{Y}_{8}:$  ...663.....662.663.....662.663.....662.663.....662.663.....662.663.....662.663.....662.663.....662.663..........
$\mathcal{Y}_{8}:$  ...2.......2...2.......2...2.......2...2.......2...2.......2...2.......2...2.......2...2.......2...2............
$\mathcal{Y}_{9}:$  ....3.1....3.1..3......3.1..3....1.3....3......3.1..3......3.1..3.1....3.1..3......3.1..3....1.3....3...........
$\mathcal{Y}_{9}:$  .....5......5..5........5..5......5....5........5..5........5....5......5..5........5..5......5....5............
$\mathcal{Y}_{9}:$  ....3.3..5.3..1.3.5..5.3..1.3.5..3.3..1.3.5..5.3..1.3.5..5.3....3.3..5.3..1.3.5..5.3..1.3.5..3.3..1.3.5..3......
$\mathcal{Y}_{9}:$  ....3..44...3...3..44...3...3..44...3...3..44...3...3..44...3...3..44...3...3..44...3...3..44...3...3..44.......
$\mathcal{Y}_{9}:$  ....3..55...3...3..55...3...3..55...3...3..55...3...3..55...3...3..55...3...3..55...3...3..55...3...3..55.......
$\mathcal{Y}_{9}:$  ..266.5..5366.266.5..5366.266.5..5366.266.5..5366.266.5..5366.266.5..5366.266.5..5366.266.5..5366.266.5..5......
$\mathcal{Y}_{9}:$  ...663.....662.663.....662.663.....662.663.....662.663.....662.663.....662.663.....662.663.....662.663..........
$\mathcal{Y}_{9}:$  ...2.......2...2.......2...2.......2...2.......2...2.......2...2.......2...2.......2...2.......2...2............
$\mathbf{.}$                                                              $\mathbf{.}$
$\mathbf{.}$                                                              $\mathbf{.}$
$\mathbf{.}$                                                              $\mathbf{.}$
$\mathbf{.}$                                                              $\mathbf{.}$
$\mathcal{Y}_{2}:$ ....3......3.1..3......3.1..3.1..1.3....3.1..1.3....3...........................................................
$\mathcal{Y}_{2}:$ ...5........5..5........5....5....5......5....5....5............................................................
$\mathcal{Y}_{2}:$ ..1.3.5..5.3..1.3.5..5.3....3.3..3.3....3.3..3.3..1.3.5..3......................................................
$\mathcal{Y}_{2}:$ ....3..44...3...3..44...3...3..44...3...3..44...3...3..44.......................................................
$\mathcal{Y}_{2}:$ ....3..55...3...3..55...3...3..55...3...3..55...3...3..55.......................................................
$\mathcal{Y}_{2}:$ ..266.5..5366.266.5..5366.266.5..5366.266.5..5366.266.5..5......................................................
$\mathcal{Y}_{2}:$ ...663.....662.663.....662.663.....662.663.....662.663..........................................................
$\mathcal{Y}_{2}:$ ...2.......2...2.......2...2.......2...2.......2...2............................................................
$\mathcal{Y}_{18}:$ ....3....1.3....3......3.1..3.1....3.1..3.......................................................................
$\mathcal{Y}_{18}:$ ...5......5....5........5....5......5..5........................................................................
$\mathcal{Y}_{18}:$ ..1.3.5..3.3..1.3.5..5.3....3.3..5.3..1.3.5..3..................................................................
$\mathcal{Y}_{18}:$ ....3..44...3...3..44...3...3..44...3...3..44...................................................................
$\mathcal{Y}_{18}:$ ....3..55...3...3..55...3...3..55...3...3..55...................................................................
$\mathcal{Y}_{18}:$ ..266.5..5366.266.5..5366.266.5..5366.266.5..5..................................................................
$\mathcal{Y}_{18}:$ ...663.....662.663.....662.663.....662.663......................................................................
$\mathcal{Y}_{18}:$ ...2.......2...2.......2...2.......2...2........................................................................
$\mathcal{Y}_{1}:$ ....3.1..1.3....3.1..1.3....3...................................................................................
$\mathcal{Y}_{1}:$ .....5....5......5....5....5....................................................................................
$\mathcal{Y}_{1}:$ ....3.3..3.3....3.3..3.3..1.3.5..3..............................................................................
$\mathcal{Y}_{1}:$ ....3..44...3...3..44...3...3..44...............................................................................
$\mathcal{Y}_{1}:$ ....3..55...3...3..55...3...3..55...............................................................................
$\mathcal{Y}_{1}:$ ..266.5..5366.266.5..5366.266.5..5..............................................................................
$\mathcal{Y}_{1}:$ ...663.....662.663.....662.663..................................................................................
$\mathcal{Y}_{1}:$ ...2.......2...2.......2...2....................................................................................
    \end{lstlisting}
    \endgroup
    \caption{Example of the queens problem on a $4 \times\ 4$ chessboard: RAAA corresponding to subsets obtained in Fig. \ref{queens_RAAA_bin_assign_00}. The ground state of this pattern is obtained in Fig. \ref{fig:queens_RAAA_activation}. For the sake of brevity, not all lines of the RAAA are presented. }
    \label{fig:queens_RAAA}
\end{figure}

\begin{figure}[htbp]
    \centering
    \fontsize{6pt}{7pt}\selectfont    
    \begin{lstlisting}   
$\mathcal{Y}_{4}$ ...0.......0...1.......0...1.......0...0.......0...0.......0...0.......1...0.......1...0.......0...1.......1
$\mathcal{Y}_{4}$ ....3......3.1..0......3.1..0......3.1..3......3.1..3......3.1..3......0.0..3.1..0.0....3.1..1.3....0.......
$\mathcal{Y}_{4}$ ...0........0..5........0..5........0..0........0..0........0..0........5....0....5......0....0....5........
$\mathcal{Y}_{4}$ ..1.3.5..5.3..0.0.5..5.3..0.0.5..5.3..1.3.5..5.3..1.3.5..5.3..1.3.5..5.0....3.3..0.0....3.3..3.3..0.0.5..3..
$\mathcal{Y}_{4}$ ....0..00...0...3..00...0...3..00...0...0..00...0...0..00...0...0..00...3...0..04...3...0..00...0...3..00...
$\mathcal{Y}_{4}$ ....3..05...3...0..05...3...0..05...3...3..05...3...3..05...3...3..05...0...3..00...0...3..50...3...0..50...
$\mathcal{Y}_{4}$ ..200.5..0300.206.5..0300.206.5..0300.200.5..0300.200.5..0300.200.5..0306.200.5..5060.000.0..5000.060.0..5..
$\mathcal{Y}_{4}$ ...060.....060.000.....060.000.....060.060.....060.060.....060.060.....000.060.....002.603.....602.003......
$\mathcal{Y}_{4}$ ...0.......0...2.......0...2.......0...0.......0...0.......0...0.......2...0.......2...0.......0...2........
$\mathcal{Y}_{8}$ ....3......3.1..0....1.3....0......3.1..3....1.3....3......3.1..3....0.0....3......0.0..3....1.3....0.......
$\mathcal{Y}_{8}$ ...0........0..5......0....5........0..0......0....0........0..0......5....0........5..0......0....5........
$\mathcal{Y}_{8}$ ..1.3.5..5.3..0.0.5..5.3..0.0.5..5.3..1.3.5..3.3..1.3.5..5.3..1.3.5..0.0..1.3.5..5.0..1.3.5..3.3..0.0.5..3..
$\mathcal{Y}_{8}$ ....0..00...0...3..00...0...3..00...0...0..00...0...0..00...0...0..04...3...0..00...3...0..00...0...3..00...
$\mathcal{Y}_{8}$ ....3..05...3...0..05...3...0..05...3...3..05...3...3..05...3...3..00...0...3..50...0...3..50...3...0..50...
$\mathcal{Y}_{8}$ ..200.5..0300.206.5..0300.206.5..0300.200.5..0300.200.5..0300.200.5..5060.000.0..5060.000.0..5000.060.0..5..
$\mathcal{Y}_{8}$ ...060.....060.000.....060.000.....060.060.....060.060.....060.060.....002.603.....002.603.....602.003......
$\mathcal{Y}_{8}$ ...0.......0...2.......0...2.......0...0.......0...0.......0...0.......2...0.......2...0.......0...2........
$\mathcal{Y}_{9}$ ....3.1....3.1..0......3.1..0....1.3....3......3.1..3......3.1..3.1....0.0..3......0.0..3....1.3....0.......
$\mathcal{Y}_{9}$ .....0......0..5........0..5......0....0........0..0........0....0......5..0........5..0......0....5........
$\mathcal{Y}_{9}$ ....3.5..5.3..0.0.5..5.3..0.0.5..3.3..1.3.5..5.3..1.3.5..5.3....3.0..5.0..1.3.5..5.0..1.3.5..3.3..0.0.5..3..
$\mathcal{Y}_{9}$ ....0..00...0...3..00...0...3..00...0...0..00...0...0..00...0...0..40...3...0..00...3...0..00...0...3..00...
$\mathcal{Y}_{9}$ ....3..05...3...0..05...3...0..05...3...3..05...3...3..05...3...3..00...0...3..50...0...3..50...3...0..50...
$\mathcal{Y}_{9}$ ..200.5..0300.206.5..0300.206.5..0300.200.5..0300.200.5..0300.200.5..5060.000.0..5060.000.0..5000.060.0..5..
$\mathcal{Y}_{9}$ ...060.....060.000.....060.000.....060.060.....060.060.....060.060.....002.603.....002.603.....602.003......
$\mathcal{Y}_{9}$ ...0.......0...2.......0...2.......0...0.......0...0.......0...0.......2...0.......2...0.......0...2........
$\mathbf{.}$                                                              $\mathbf{.}$
$\mathbf{.}$                                                              $\mathbf{.}$
$\mathbf{.}$                                                              $\mathbf{.}$
$\mathbf{.}$                                                              $\mathbf{.}$
$\mathcal{Y}_{2}:$ ....3......3.1..0......3.1..0.0..1.3....3.1..1.3....3.......................................................
$\mathcal{Y}_{2}:$ ...0........0..5........0....5....0......0....0....0........................................................
$\mathcal{Y}_{2}:$ ..1.3.5..5.3..0.0.5..5.3....0.0..3.3....3.3..3.3..1.3.5..3..................................................
$\mathcal{Y}_{2}:$ ....0..00...0...3..00...0...3..40...0...0..00...0...0..00...................................................
$\mathcal{Y}_{2}:$ ....3..05...3...0..05...3...0..00...3...3..50...3...3..50...................................................
$\mathcal{Y}_{2}:$ ..200.5..0300.206.5..0300.206.5..5000.000.0..5000.000.0..5..................................................
$\mathcal{Y}_{2}:$ ...060.....060.000.....060.000.....602.603.....602.603......................................................
$\mathcal{Y}_{2}:$ ...0.......0...2.......0...2.......0...0.......0...0........................................................
$\mathcal{Y}_{18}:$ ....3....1.3....0......3.1..0.0....0.0..0...................................................................
$\mathcal{Y}_{18}:$ ...0......0....5........0....5......5..5....................................................................
$\mathcal{Y}_{18}:$ ..1.3.5..3.3..0.0.5..5.3....0.0..5.0..0.0.5..3..............................................................
$\mathcal{Y}_{18}:$ ....0..00...0...3..00...0...3..40...3...3..00...............................................................
$\mathcal{Y}_{18}:$ ....3..05...3...0..05...3...0..00...0...0..50...............................................................
$\mathcal{Y}_{18}:$ ..200.5..0300.206.5..0300.206.5..5060.060.0..5..............................................................
$\mathcal{Y}_{18}:$ ...060.....060.000.....060.000.....002.003..................................................................
$\mathcal{Y}_{18}:$ ...0.......0...2.......0...2.......2...2....................................................................
$\mathcal{Y}_{1}:$ ....3.1..1.3....0.0..1.3....0...............................................................................
$\mathcal{Y}_{1}:$ .....0....0......5....0....5................................................................................
$\mathcal{Y}_{1}:$ ....3.3..3.3....0.0..3.3..0.0.5..3..........................................................................
$\mathcal{Y}_{1}:$ ....0..00...0...3..40...0...3..00...........................................................................
$\mathcal{Y}_{1}:$ ....3..05...3...0..00...3...0..50...........................................................................
$\mathcal{Y}_{1}:$ ..200.5..0300.206.5..5000.060.0..5..........................................................................
$\mathcal{Y}_{1}:$ ...060.....060.000.....602.003..............................................................................
$\mathcal{Y}_{1}:$ ...0.......0...2.......0...2................................................................................
    \end{lstlisting}
    \caption{Example of the queens problem on a $4 \times\ 4$ chessboard: The corresponding RAA configuration yields an activation pattern whose ground state encodes a valid solution to the queens problem. The active decision variables in this solution are \(q_{0,2}, q_{1,0}, q_{2,3},\) and \(q_{3,1}\), corresponding to non-attacking queen placements. For the sake of brevity, not all lines of the RAAA are presented. }
    \label{fig:queens_RAAA_activation}
\end{figure}

\end{widetext}

\clearpage

\bibliography{ConstrGAP_MWIS_Graph}
\end{document}